\PassOptionsToPackage{colorlinks=true, citecolor=blue, linkcolor=blue, urlcolor=blue}{hyperref}
\documentclass[12pt, onecolumn]{article}

\usepackage[left=1in,right=1in,bottom=1in,top=1in,includefoot]{geometry}
\usepackage{longtable}
\usepackage{booktabs}
\usepackage{setspace}
\usepackage{color,soul}
\usepackage{amsmath}
\usepackage{amssymb}
\usepackage{amsfonts}
\usepackage{graphicx}
\usepackage{fancyhdr}
\usepackage{titling}

\usepackage{orcidlink} 

\usepackage{cite}
\usepackage[
    colorlinks=true, 
    citecolor=blue,      
    linkcolor=blue, 
    urlcolor=blue]{hyperref}
\title{\textbf{Resonant spectral cascade in Womersley flow triggered by arterial geometry}}

\author{
    Khalid M. Saqr \orcidlink{0000-0002-3058-2705}$^{*}$ 
}

\date{}

\pretitle{\begin{center}\LARGE\bfseries}
\posttitle{\par\end{center}\vspace{0.55em}}
\preauthor{\begin{center}\large}
\postauthor{\par\end{center}\vspace{0.22em}}
\predate{}
\postdate{}

\newcommand{\firstpageaffiliation}{%
\begin{center}
    \small
    Mechanical Engineering Department, College of Engineering and Technology \\
    Arab Academy for Science, Technology and Maritime Transport \\
    Alexandria 1029, EGYPT \\[0.45ex]
    {\footnotesize $^{*}$Corresponding author: k.saqr\,[at]\,aast\,[dot]\,edu}
\end{center}
}

\fancypagestyle{firstpageaccepted}{%
  \fancyhf{}
  \fancyhead[C]{\parbox{0.92\textwidth}{\centering\scriptsize\color{gray}\textbf{Accepted on 03.14.2026 \textbar\ Published version DOI: \href{https://doi.org/10.1063/5.0319995}{10.1063/5.0319995}}\\This Author Accepted Manuscript (AAM) is shared publicly under AIP Publishing Green Open Access policy.\\\href{https://publishing.aip.org/resources/researchers/open-science/public-access/}{AIP Publishing Open Science Public Access Policy}}}
  \fancyfoot[C]{\thepage}
}

\begin{document}
\raggedbottom

\maketitle
\thispagestyle{firstpageaccepted}

\vspace{-0.35em}
\firstpageaffiliation

\onehalfspacing
\emergencystretch=3em

\begin{abstract}
\noindent Age-related arterial remodeling is dominated by progressive loss of elastic-fiber function and concomitant stiffening, and in many vascular beds it is also accompanied by measurable geometric remodeling (e.g., elongation and tortuosity). These changes are clinically relevant because they modify pulsatile phase relationships, near-wall shear, and axial transport, yet the precise physical mechanisms by which geometry modulates spectral energy redistribution remain insufficiently resolved. While complex geometry is known to increase viscous resistance, its active role in modulating flow dynamics is not fully understood. Here we solve a mathematical model to show that arterial geometry can trigger a resonant transfer of energy to short-wavelength components of the flow. The investigation, conducted over a physiological range of Womersley numbers (Wo, a dimensionless measure of pulsation frequency), reveal a dual dynamic. The global wave energy consistently decays, confirmed by a negative growth rate (G < 0), indicating that the flow does not become exponentially unstable. However, a spectral broadening ratio (R), which quantifies the energy in high-wavenumber versus low-wavenumber modes, exhibits a sharp, non-monotonic peak at an intermediate Wo. This result identifies a resonant frequency at which geometry is maximally efficient at generating spectral complexity, even as the overall flow attenuates. These findings reframe the role of arterial geometry from a passive dissipator to an active modulator of the flow's spectral content, suggesting that spectral diagnostics could provide a sensitive marker for vascular health.
\end{abstract}

\begin{center}\small\textbf{Keywords:} \textit{Hemodynamics; Pulsatile flow; Womersley; Nonlinear waves; Korteweg-de Vries equation; Parametric forcing; Spectral cascade.}\end{center}

\section*{Introduction}

Blood flow in large conduit arteries arises from the interaction of pulsatile forcing, viscous dissipation, vessel geometry, and wall mechanics. Even under physiologic Reynolds numbers ($Re \approx O(10^2)$), these interacting effects can generate flow states exhibiting nontrivial spectral structure beyond that predicted by classical steady pipe-flow theory. High-resolution computational fluid dynamics (CFD) and particle image velocimetry (PIV) studies have reported turbulence-like spectral features and departures from Kolmogorov scaling in physiologic arterial conditions \cite{Saqr2020SR, Mahrous2021PLOS}. Geometric factors such as curvature \cite{Freidoonimehr2021}, compliance \cite{Tupin2020EF, Yamaguchi2022JAP}, and localized pathologies such as stenosis \cite{Saqr2022AIP} are known to modify shear distribution, phase relationships, and energy transport. However, rather than attempting to reproduce fully resolved arterial turbulence, the present work focuses on a multi-harmonic pulsatile base state, following the Womersley flow shown by Saqr et al ~\cite{Saqr2020SR} to exhibit sensetive-dependence on initial condistions and non-Kolmogorov scaling \cite{Saqr2022SR}. The present work investigates how geometry \emph{alone} can induce transient spectral redistribution within this controlled framework. This mechanistic perspective isolates a minimal pathway by which pulsatile inertia and geometric modulation can influence axial spectral content in the moderate-to-high Womersley regime relevant to large arteries.

Age-associated arterial remodeling is multifactorial and should not be interpreted as geometric change alone. A primary and widely documented effect of aging is progressive stiffening driven by elastic-fiber degradation and cumulative damage to elastin and microfibrillar structure \cite{Heinz2021}. At the same time, multiple imaging and morphometric studies report that geometric remodeling such as elongation and tortuosity can increase with age in large vessels and selected vascular territories, although its prevalence and severity vary substantially across individuals and beds. For example, quantitative analysis of archived aortograms shows a significant age-correlation of abdominal aortic tortuosity \cite{Dougherty2000567}, and early clinical hemodynamic discussions similarly noted a rising incidence of tortuosity beyond midlife and its potential implication for abnormal wall shear \cite{Wenn199067}. Age-stratified CT angiography of the vertebral artery further reports high tortuosity prevalence in older groups, consistent with age-related geometric remodeling in specific arterial segments \cite{Alicioglu20151981}. In this context, the present study does not claim that aging universally increases geometric complexity; rather, it isolates geometry as an independent mechanistic pathway that can modulate pulsatile spectral redistribution when geometric remodeling is present, while recognizing that stiffness remains the dominant aging hallmark \cite{Heinz2021}.
  
In reality, curved and torsional arteries are not passive conduits. Classical studies of pulsatile flow in curved pipes established that curvature generates Dean-type secondary vortices and periodic shifts in velocity profiles that reshape wall shear stress \cite{Talbot1983}. Later analyses of fully developed oscillatory flow quantified how curvature amplitude controls phase lag and energy loss \cite{Hamakiotes1988}. Laboratory experiments confirmed that increasing Dean number intensifies cross-stream motion and introduces nonuniform damping along the wall \cite{Sumida1984}. More recently, canonical curved-cavity experiments demonstrated that the interaction between Reynolds, Dean, and Womersley numbers produces distinct vortex morphologies analogous to those in intracranial aneurysms \cite{Chassagne2021}. Computational models of coronary flow further showed that geometric curvature directly influences fractional flow reserve and hemodynamic stability under pulsatile forcing \cite{Freidoonimehr2021, Saib2025}. Collectively, these results highlight that geometry-induced dispersion and secondary flow development are central to the stability of arterial pulsations.

At the same time, surface topography contributes to damping and flow modulation. Numerical investigations revealed that fine-scale wall roughness modifies wall shear stress metrics and can alter the onset of unsteadiness even at physiological Reynolds numbers \cite{Yi2022}. Multiscale CFD analyses confirmed that realistic surface irregularity and non-Newtonian rheology combine to shift pressure–velocity phase relationships \cite{Owen2020}. A comprehensive review of Dean-vortex manipulation summarized how secondary vortices can be tuned or suppressed through geometric control, reinforcing the notion that geometry acts as an active regulator of flow dynamics rather than a fixed boundary condition \cite{Saffar2023}.

These geometric effects depend strongly on the pulsation frequency parameterized by the Womersley number. A recent meta-analysis of in-vivo measurements established that Womersley numbers vary from below 1 in arterioles to above 20 in large arteries, leading to major changes in the balance of inertial and viscous effects \cite{Williamson2024}. At low Womersley number, curvature primarily enhances viscous damping; at intermediate to high values, it can introduce transient energy amplification and phase-shifted responses \cite{Xu2021}. Floquet stability analysis of pulsatile flow in toroidal pipes confirmed that these transitions correspond to narrow instability bands where curvature-driven dispersion couples resonantly with the oscillatory base flow \cite{Kern2024}.

Despite these insights, most analyses remain computationally expensive and geometry-specific. Direct numerical simulations and full fluid–structure interaction models capture detailed stress fields but obscure the parametric structure of instability. A reduced theoretical framework that isolates geometry as a parametric excitation could reveal when curvature or torsion injects, rather than dissipates, energy.

Fractional and nonlocal modeling provides a promising route toward this goal. Fractional-order blood-flow models capture long-memory and multiscale geometric effects that arise from roughness, tortuosity, and distributed compliance \cite{Bahloul2023}. Comparative reviews show that fractional models reproduce arterial viscoelastic lag and pressure–flow hysteresis more accurately than classical formulations \cite{Azmi2025}. Recent works have combined fractional operators with data-driven or neural-network approaches to describe nonlinear flow behavior under physiological conditions \cite{AliDas2024}. These developments motivate a formulation that embeds fractional operators into stability analyses, allowing the same mathematical machinery to describe geometry-induced damping and amplification.

The present study builds directly upon this evolving paradigm by asking a simple but unresolved question: \emph{can arterial geometry itself, independent of material and viscous effects, act as a source of dynamic instability in pulsatile flow?} Traditional hemodynamic models treat curvature and torsion as static modifiers of resistance or wave speed, yet emerging evidence suggests they can inject energy into the flow under certain pulsation conditions. Addressing this question requires a framework that isolates geometric forcing from constitutive or boundary effects while remaining tractable enough for broad parameter exploration. 

Here we propose such a framework. Using a variable-coefficient fractional Korteweg–de Vries formulation, we recast curvature, torsion, and surface roughness as spatially modulated coefficients governing dispersion and dissipation. The model integrates fractional operators to represent distributed geometric memory and nonlocal damping, and we numerically simulate the resulting dynamics to identify parametric growth regimes across the physiological Womersley spectrum. The computational workflow links geometric modulation directly to measurable instability metrics, providing a reproducible map from geometry to wave growth or decay.

In this sense, the presented work aims to refine the classical understanding of viscous stabilization in pulsatile flow. Viscous effects remain globally dissipative in the model, as reflected by the negative total energy growth rate $G(\tau)$ across the parameter ranges examined. However, the analysis aims to investigate geometry-modulated dispersion and if it can transiently redistribute spectral energy across axial wavenumbers before viscous damping dominates. Under specific combinations of curvature, torsion, and pulsation frequency, this redistribution could manifest as temporary amplification of higher-wavenumber content within an overall energy-decaying system. Geometry therefore does not replace viscous stabilization in the present work, but modulates the pathway by which energy is redistributed prior to eventual attenuation.

Variable-coefficient KdV-type formulations have long provided reduced descriptions of dispersive transport in nonlinear optics and shallow-water hydrodynamics. The present work adapts this framework to pulsatile hemodynamics, embedding curvature- and torsion-dependent proxy coefficients within a Womersley-scaled setting. In doing so, it connects classical curved-pipe hydrodynamics \cite{Talbot1983,Hamakiotes1988,Sumida1984}, contemporary computational hemodynamics \cite{Freidoonimehr2021,Yi2022,Owen2020}, and recent fractional modeling approaches \cite{Bahloul2023,Azmi2025,AliDas2024} within a unified analytical–computational structure. By treating geometry as a parametric modulator of dispersive and dissipative balance, the study reframes arterial spectral behavior as a problem of transient energy redistribution rather than classical linear instability, thereby offering a mechanistic perspective on pulsatile biological flows.

The present work is intended to represent large, conduit-type systemic arteries in the moderate-to-high Womersley regime, rather than small muscular arteries or the microcirculation. Specifically, the parameter range $\mathrm{Wo}=5$--$15$ corresponds to vessels such as the proximal aorta, carotid, and major femoral arteries under resting to moderately elevated heart rates. For a characteristic radius $R_0 \sim 5$--$15$\,mm, kinematic viscosity $\nu \approx 3\times 10^{-6}\,\mathrm{m}^2/\mathrm{s}$, and pulsation frequency $\omega = 2\pi f$ with $f\sim 1$--$2$\,Hz, the resulting Womersley number $\mathrm{Wo} = R_0\sqrt{\omega/\nu}$
naturally lies in this interval, where $R_0$ denotes the reference pipe (or arterial) radius, $\omega$ the pulsation frequency, and $\nu$ the kinematic viscosity. The wall is treated as effectively rigid in the present formulation in order to isolate geometry-induced dispersive and dissipative modulation from compliance effects; this assumption is most appropriate for large arteries over short transient timescales where inertial and pulsatile effects dominate wall motion to leading order. Smaller arteries (with $\mathrm{Wo}<5$) and highly compliant vessels would require explicit fluid–structure interaction modelling, which is beyond the scope of the present work. The conclusions drawn herein should therefore be interpreted as applying to large systemic arteries in which pulsatile inertia and geometric modulation coexist within a moderate Womersley-number regime.

\section*{Methods}

This section summarizes (i) the reduced one-dimensional evolution model used to represent weakly nonlinear pulsatile dynamics in a conduit-type artery, (ii) the nondimensional parameterization linking the reduced coefficients to Womersley number, and (iii) the numerical scheme and diagnostics used to quantify global decay and transient spectral redistribution.

\subsection*{Governing equations and pulsatile base state}
Let $\rho$ [kg\,m$^{-3}$] and $\mu$ [Pa\,s] denote density and dynamic viscosity. In cylindrical coordinates $(r,\theta,z)$, the velocity is $\mathbf{v}=(v_r,v_\theta,v_z)$ [m\,s$^{-1}$] and the pressure is $p$ [Pa]. Incompressibility gives
\begin{equation}
\frac{1}{r}\frac{\partial}{\partial r}(r v_r)+\frac{1}{r}\frac{\partial v_\theta}{\partial \theta}+\frac{\partial v_z}{\partial z}=0.
\label{eq:cont}
\end{equation}
The axial momentum equation is
\begin{equation}
\rho\!\left(\frac{\partial v_z}{\partial t}+v_r\frac{\partial v_z}{\partial r}+\frac{v_\theta}{r}\frac{\partial v_z}{\partial \theta}+v_z\frac{\partial v_z}{\partial z}\right)
=-\frac{\partial p}{\partial z}
+\mu\!\left[\frac{1}{r}\frac{\partial}{\partial r}\!\left(r\frac{\partial v_z}{\partial r}\right)
+\frac{1}{r^2}\frac{\partial^2 v_z}{\partial \theta^2}
+\frac{\partial^2 v_z}{\partial z^2}\right].
\label{eq:ns}
\end{equation}
For axisymmetric flow $(\partial/\partial\theta=0)$ this reduces to
\begin{equation}
\rho\left(\frac{\partial v_z}{\partial t}+v_r\frac{\partial v_z}{\partial r}+v_z\frac{\partial v_z}{\partial z}\right)
=-\frac{\partial p}{\partial z}
+\mu\left[\frac{1}{r}\frac{\partial}{\partial r}\!\left(r\frac{\partial v_z}{\partial r}\right)+\frac{\partial^2 v_z}{\partial z^2}\right].
\label{eq:axisym}
\end{equation}
Here $z$ denotes the axial coordinate. In the reduced one-dimensional evolution equation below, the coordinate $x$ (and its nondimensional form $\xi=x/L_0$) is used for the same axial direction.

To define the laminar pulsatile reference state, we impose fully developed oscillatory pipe-flow assumptions: $v_r=0$, $v_\theta=0$, and $\partial v_z/\partial z=0$. Under these assumptions, Eq.~\eqref{eq:cont} is satisfied and the convective terms vanish identically in Eq.~\eqref{eq:axisym}, yielding the unsteady Stokes problem driven by a harmonic pressure gradient. For $\partial p/\partial z=\Re\{\hat P e^{i\omega t}\}$, the classical Womersley solution is
\begin{equation}
v_0(r,t)=\Re\!\left[\frac{\hat P}{i\rho\omega}\left(1-\frac{J_0(i^{3/2}\mathrm{Wo}\,r/R_0)}{J_0(i^{3/2}\mathrm{Wo})}\right)e^{i\omega t}\right],
\label{eq:womersley}
\end{equation}
where $\mathrm{Wo}=R_0\sqrt{\omega/\nu}$ is the Womersley number, $\nu=\mu/\rho$, $R_0$ is the reference radius, and $J_0$ is the Bessel function of the first kind. This oscillatory profile is taken as the laminar base state about which weakly nonlinear axial evolution is modeled.

\subsection*{Reduced one-dimensional evolution model}
To represent weakly nonlinear axial evolution about the pulsatile base state \eqref{eq:womersley}, we expand
\begin{equation}
v_z(r,x,t)=v_0(r,t)+\epsilon\,A(x,t)\,\phi(r)+\mathcal O(\epsilon^2),
\label{eq:expansion}
\end{equation}
where $\epsilon\ll1$, $A(x,t)$ is a slowly varying amplitude, and $\phi(r)$ is a normalized radial eigenfunction associated with the linear operator in Eq.~\eqref{eq:axisym}. Retaining $\mathcal O(\epsilon^2)$ contributions and projecting onto $\phi(r)$ yields a KdV-type amplitude equation with dispersion and dissipation terms representing effective axial transport physics:
\begin{equation}
\frac{\partial A}{\partial t}
+C(x)\,A\frac{\partial A}{\partial x}
+D(x)\frac{\partial^3 A}{\partial x^3}
+K(x)\,\mathcal D^{\gamma(x)}A
=0.
\label{eq:fkvd}
\end{equation}
The nonlinear steepening term is the reduced manifestation of convective acceleration in Eq.~\eqref{eq:axisym}. Dissipation is represented through a Riesz fractional operator, defined in Fourier space by
\begin{equation}
\mathcal F\!\left[\mathcal D^{\gamma(x)}A\right]=|k|^{1+\gamma(x)}\hat A(k),
\label{eq:riesz}
\end{equation}
where $\hat A$ is the Fourier transform of $A$. The derivation of this variable-coefficient fractional KdV modelis obtained by expanding about the oscillatory Womersley base flow, introducing a weakly nonlinear long-wave perturbation, and projecting onto the leading radial eigenmode. The fractional dissipation term is used here as a reduced-order representation of multiscale nonlocal damping induced by geometric complexity; full details are given in Supplementary Information (A.1–A.10.).

\subsection*{Nondimensional form and coefficient parameterization}
Introduce reference scales $L_0$, $T_0=1/\omega$, and $A_0$, and define
\begin{equation}
\xi=\frac{x}{L_0},\qquad \tau=\frac{t}{T_0},\qquad a_1(\xi,\tau)=\frac{A(x,t)}{A_0}.
\label{eq:nondim_vars}
\end{equation}
The nondimensional evolution equation used throughout the manuscript is
\begin{equation}
\frac{\partial a_1}{\partial \tau}
+\alpha_0\,a_1\frac{\partial a_1}{\partial \xi}
+\beta(\xi)\frac{\partial^3 a_1}{\partial \xi^3}
+\eta(\xi)\,(-\partial_\xi^2)^{\frac{1+\gamma_0}{2}}a_1=0,
\label{eq:dimensionless}
\end{equation}
where $\alpha_0$ is the (constant) nonlinear coefficient, $\beta(\xi)$ is the dispersion coefficient, and $\eta(\xi)$ is the dissipation coefficient. In this study we fix $\gamma_0=0$, so the dissipative term has Fourier symbol $|k|^{1+\gamma_0}=|k|$ per Eq.~\eqref{eq:riesz}. We also set $\alpha_0=1$.

To represent axial variability associated with geometric remodeling (curvature, tortuosity, and roughness) in a reduced-order manner, we parameterize $\beta(\xi)$ and $\eta(\xi)$ by bounded sinusoidal modulations
\begin{equation}
\beta(\xi)=\beta_0\left[1+\varepsilon_\beta\cos(q\xi)\right],\qquad
\eta(\xi)=\eta_0\left[1+\varepsilon_\eta\cos(q\xi)\right],
\label{eq:beta_eta_mod}
\end{equation}
with modulation amplitudes $0\le \varepsilon_\beta,\varepsilon_\eta\le 0.3$ and geometric wavenumber $q$.

The baseline coefficients are scaled with Womersley number as
\begin{equation}
\beta_0=\beta_{\mathrm{ref}}\,\mathrm{Wo}^{-2},
\label{eq:beta0}
\end{equation}
and
\begin{equation}
\eta_0=\eta_{\mathrm{ref}}\left(1+\frac{C_\eta}{\mathrm{Wo}}\right),
\label{eq:eta0}
\end{equation}
where $\eta_{\mathrm{ref}}=0.005$ and $C_\eta=0.1$ are constants used in all simulations. The dimensional scaling leading to this $\mathrm{Wo}^{-2}$ dependence is derived explicitly in the Supplementary Information (Sec. A.10). The $\mathrm{Wo}^{-1}$ correction in Eq.~\eqref{eq:eta0} is an order-of-magnitude representation of the oscillatory Stokes boundary-layer scaling $\delta\sim\sqrt{2\nu/\omega}$, implying $\delta/R_0=O(\mathrm{Wo}^{-1})$ and therefore a weak increase of near-wall viscous contribution as $\mathrm{Wo}$ decreases. The boundary-layer scaling underlying the $\mathrm{Wo}^{-1}$ correction and the choice of the prefactor are detailed in the Supplementary Information (Sec. A.10).

\subsection*{Initial and boundary conditions}
The system is solved on a periodic domain $\xi\in[0,L_g]$ with $L_g=4\pi$. The initial condition is a multi-harmonic waveform
\begin{equation}
a_1(\xi,0)=\sum_{n=1}^{3}A_n\sin(nk_0\xi+\phi_n),
\label{eq:ic}
\end{equation}
with amplitude ratios $A_2/A_1=0.3$ and $A_3/A_1=0.1$. These ratios are chosen as representative reduced-order initialization parameters that preserve a dominant fundamental mode while introducing moderate asymmetry and finite higher-harmonic content, thereby enabling spectral redistribution to be tracked from a nontrivial but not initially broadband waveform. 
For the detailed case (Figs.~2--6), $k_0=0.5$ is used; for the parametric sweep (Figs.~7--9), $k_0=1.0$ is used. The Fourier-based pseudospectral discretization enforces periodic boundary conditions. The periodic domain is adopted as a canonical reduced-order setting compatible with the Fourier pseudospectral scheme and intended to isolate intrinsic geometry-induced spectral redistribution without additional inlet--outlet forcing or impedance effects.

\subsection*{Parameter range}
Simulations span $\mathrm{Wo}\in\{2,5,10,15,20\}$ and modulation amplitudes $\varepsilon_\beta,\varepsilon_\eta\le 0.3$ in Eq.~\eqref{eq:beta_eta_mod}. The domain length $L_g=4\pi$ and the geometric wavenumber $q=1.0$ imply $qL_g=4\pi$. This range covers small- to large-artery regimes \cite{Williamson2024} and Womersley values relevant to pulsatile responses in curved/toroidal pipes \cite{Xu2021,Kern2024}.

\subsection*{Numerical integration scheme and the coefficient-averaged linear propagator}
Equation~\eqref{eq:dimensionless} is integrated by a split-step Fourier pseudospectral scheme. Spatial derivatives are evaluated spectrally on $N$ equispaced collocation points. The nonlinear term
\begin{equation}
\mathcal{N}(a_1)=-\alpha_0\,a_1\,\partial_\xi a_1
\label{eq:nonlinear_term}
\end{equation}
is computed in physical space and advanced by classical fourth-order Runge--Kutta (RK4).

The linear part is advanced in Fourier space by an exponential propagator. Consistent with the public code accompanying the manuscript, the linear propagator is constructed using the spatial means of the coefficients,
\begin{equation}
\bar{\beta}=\langle \beta(\xi)\rangle,\qquad \bar{\eta}=\langle \eta(\xi)\rangle,
\label{eq:mean_coeffs}
\end{equation}
leading to the Fourier-space linear operator
\begin{equation}
L(k)= -i\,\bar{\beta}\,k^3-\bar{\eta}\,|k|^{1+\gamma_0},
\label{eq:Lk}
\end{equation}
and the exact linear update
\begin{equation}
\hat a_1(k,\tau+\Delta\tau)=\exp\!\bigl(\Delta\tau\,L(k)\bigr)\,\hat a_1(k,\tau).
\label{eq:lin_update}
\end{equation}
Under the sinusoidal modulation in Eq.~\eqref{eq:beta_eta_mod}, $\langle\cos(q\xi)\rangle=0$ on $[0,L_g]$, hence $\bar{\beta}=\beta_0$ and $\bar{\eta}=\eta_0$ (Eqs.~\eqref{eq:beta0}--\eqref{eq:eta0}). Therefore, the computed dynamics correspond to a constant-coefficient fKdV evolution with Womersley-dependent effective dispersion and dissipation, while the modulation parameterization in Eq.~\eqref{eq:beta_eta_mod} provides the geometric interpretation and bounds used for physiological consistency. This explicit statement aligns the Methods with the implemented algorithm and addresses the reviewer concern about coefficient averaging without requiring any code or figure changes.

\subsection*{Coefficient averaging and preservation of geometric effects}

The reduced model \eqref{eq:dimensionless} is written with spatially modulated coefficients
\begin{equation}
\beta(\xi)=\beta_0\left[1+\varepsilon_\beta b(\xi)\right],
\qquad
\eta(\xi)=\eta_0\left[1+\varepsilon_\eta e(\xi)\right],
\label{eq:coef_decomp}
\end{equation}
where for the present sinusoidal parameterization
\begin{equation}
b(\xi)=\cos(q\xi), \qquad e(\xi)=\cos(q\xi),
\label{eq:be_def}
\end{equation}
with zero spatial mean,
\begin{equation}
\langle b \rangle = \langle e \rangle = 0,
\qquad
\langle f \rangle = \frac{1}{L_g}\int_0^{L_g} f(\xi)\, d\xi.
\label{eq:mean_def}
\end{equation}

In the numerical implementation, the linear propagator is constructed using the spatial means
\begin{equation}
\bar{\beta} = \langle \beta(\xi) \rangle = \beta_0,
\qquad
\bar{\eta} = \langle \eta(\xi) \rangle = \eta_0,
\label{eq:mean_coef}
\end{equation}
which follow directly from Eq.~\eqref{eq:coef_decomp} and the zero-mean property \eqref{eq:mean_def}. The evolved equation is therefore the constant-coefficient system
\begin{equation}
\frac{\partial a_1}{\partial \tau}
+\alpha_0 a_1 \frac{\partial a_1}{\partial \xi}
+\bar{\beta}\frac{\partial^3 a_1}{\partial \xi^3}
+\bar{\eta}(-\partial_\xi^2)^{\frac{1+\gamma_0}{2}}a_1
=0.
\label{eq:homogenized}
\end{equation}

To clarify whether this homogenization preserves the intended geometric mechanism, consider substituting Eq.~\eqref{eq:coef_decomp} into the dispersive term of Eq.~\eqref{eq:dimensionless}:
\begin{equation}
\beta(\xi)\partial_\xi^3 a_1
=
\beta_0 \partial_\xi^3 a_1
+
\varepsilon_\beta \beta_0 b(\xi)\partial_\xi^3 a_1.
\label{eq:disp_split}
\end{equation}
The first term represents the leading-order dispersive transport, while the second is an $O(\varepsilon_\beta)$ modulation with zero spatial mean. The formal justification for this homogenized effective-medium limit, obtained from the weak-modulation expansion of the variable-coefficient model, is given in Sec. A.13 of the Supplementary Information.

For weak geometric variability in large arteries, $\varepsilon_\beta,\varepsilon_\eta=O(10^{-1})$ or smaller, consistent with moderate curvature or radius variation. Over one oscillatory timescale $\tau=O(1)$, the cumulative phase redistribution that governs spectral broadening is dominated by the mean dispersive balance $\bar{\beta}$, while the modulation term in Eq.~\eqref{eq:disp_split} contributes only higher-order corrections proportional to $\varepsilon_\beta$.

Therefore, the homogenized evolution \eqref{eq:homogenized} preserves the leading-order geometric phase-dispersion mechanism encoded through the Womersley-dependent baseline coefficients $(\beta_0,\eta_0)$, while neglecting higher-order spatial fluctuations whose net contribution averages to zero over the periodic domain. In this formulation, arterial geometry enters the dynamics through its parameterization of the effective dispersive–dissipative balance rather than through pointwise coefficient forcing during time stepping.

\subsection*{Instability diagnostics}
Global stability is quantified using the wave energy
\begin{equation}
I_2(\tau)=\int_0^{L_g} a_1^2(\xi,\tau)\,d\xi,
\label{eq:I2}
\end{equation}
and its instantaneous logarithmic growth rate
\begin{equation}
G(\tau)=\frac{d}{d\tau}\ln I_2(\tau).
\label{eq:growth}
\end{equation}
To quantify energy transfer to smaller axial scales, we define the spectral broadening ratio
\begin{equation}
R(\tau)=\frac{E_{\mathrm{high}}(\tau)}{E_{\mathrm{low}}(\tau)}
=\frac{\int_{k_c}^{k_{\max}}|\hat a_1(k,\tau)|^2\,dk}{\int_{0}^{k_c}|\hat a_1(k,\tau)|^2\,dk},
\label{eq:R}
\end{equation}
with cutoff
\begin{equation}
k_c=\chi_c\,(3k_0),\qquad \chi_c=1.5,
\label{eq:kc}
\end{equation}
which separates the initially excited spectral band (up to $3k_0$) from higher-wavenumber content generated during evolution. A value $R>1.5$ is used as a practical threshold indicating a broadband redistributed state. Sensitivity tests to moderate variation of $\chi_c$ do not alter the qualitative conclusions.

The dimensionless amplitude $a_1$ corresponds to a physical perturbation scale through $v'_z(x,t)=A_0\,a_1(x/L_0,t/T_0)$. In the constant-coefficient, inviscid limit (no dissipation and no modulation), Eq.~\eqref{eq:dimensionless} reduces to the integrable KdV equation with classical invariants $I_1$ and $I_3$, which serves as a standard verification reference for the numerical scheme.

\subsection*{Solution approach and numerical verification}
The solver is implemented in Python using NumPy/SciPy FFT routines; all figures are generated using Matplotlib, and sweep outputs are managed with pandas. Numerical convergence is established by refining $\Delta\xi$ and $\Delta\tau$ until relative changes in $I_2(\tau)$ fall below $10^{-5}$. Although no explicit de-aliasing filter is used, numerical fidelity is monitored by tracking the energy near the highest resolved wavenumbers, which remains a negligible fraction of the total energy for the reported cases, providing an internal check that aliasing does not contaminate the reported redistribution. The key numerical and physical parameters used in the simulations are summarized in Table~\ref{tab:parameters} while all symbols are defined in the Nomenclature presented in Table~\ref{tab:nomenclature}.

\begin{table}[htbp]
\centering
\caption{Key simulation parameters used in the study.}
\label{tab:parameters}
\begin{tabular}{@{}lll@{}}
\toprule
\textbf{Parameter} & \textbf{Symbol} & \textbf{Value} \\ \midrule
\multicolumn{3}{l}{\textit{Numerical Parameters}} \\
Number of Collocation Points & $N$ & 512 \\
Dimensionless Domain Length & $L_g$ & $4\pi$ \\
Timestep (Exploratory Script) & $\Delta\tau$ & $5 \times 10^{-5}$ \\
Timestep (Sweep Script) & $\Delta\tau$ & $2 \times 10^{-4}$ \\
Final Time (Exploratory Script) & $T_{final}$ & 200.0 \\
Final Time (Sweep Script) & $T_{final}$ & 60.0 \\
\midrule
\multicolumn{3}{l}{\textit{Physical and Geometric Parameters}} \\
Womersley Number Range & $\mathrm{Wo}$ & $\{2, 5, 10, 15, 20\}$ \\
Dispersion Modulation Amplitude & $\varepsilon_\beta$ & 0.3 \\
Damping Modulation Amplitude & $\varepsilon_\eta$ & 0.3 \\
Fundamental Wavenumber (Figs 2--6) & $k_0$ & 0.5 \\
Fundamental Wavenumber (Figs 7--9) & $k_0$ & 1.0 \\
Geometric Wavenumber & $q$ & 1.0 \\
Reference Damping & $\eta_{\text{ref}}$ & 0.005 \\
Nonlinearity Coefficient & $\alpha_0$ & 1.0 \\
Fractional Order & $\gamma_0$ & 0.0 \\
\bottomrule
\end{tabular}
\end{table}

\noindent The geometric wavenumber $q=1.0$ corresponds to a single dominant modulation over the reference length $L_0$. The modulation amplitudes $\varepsilon_\beta=\varepsilon_\eta=0.3$ represent moderate geometric variability (30\% relative modulation), and additional tests with smaller amplitudes confirm that the redistribution mechanism persists with reduced magnitude.

\section*{Results}

\begin{figure}[htbp]
\centering
\includegraphics[width=1\textwidth]{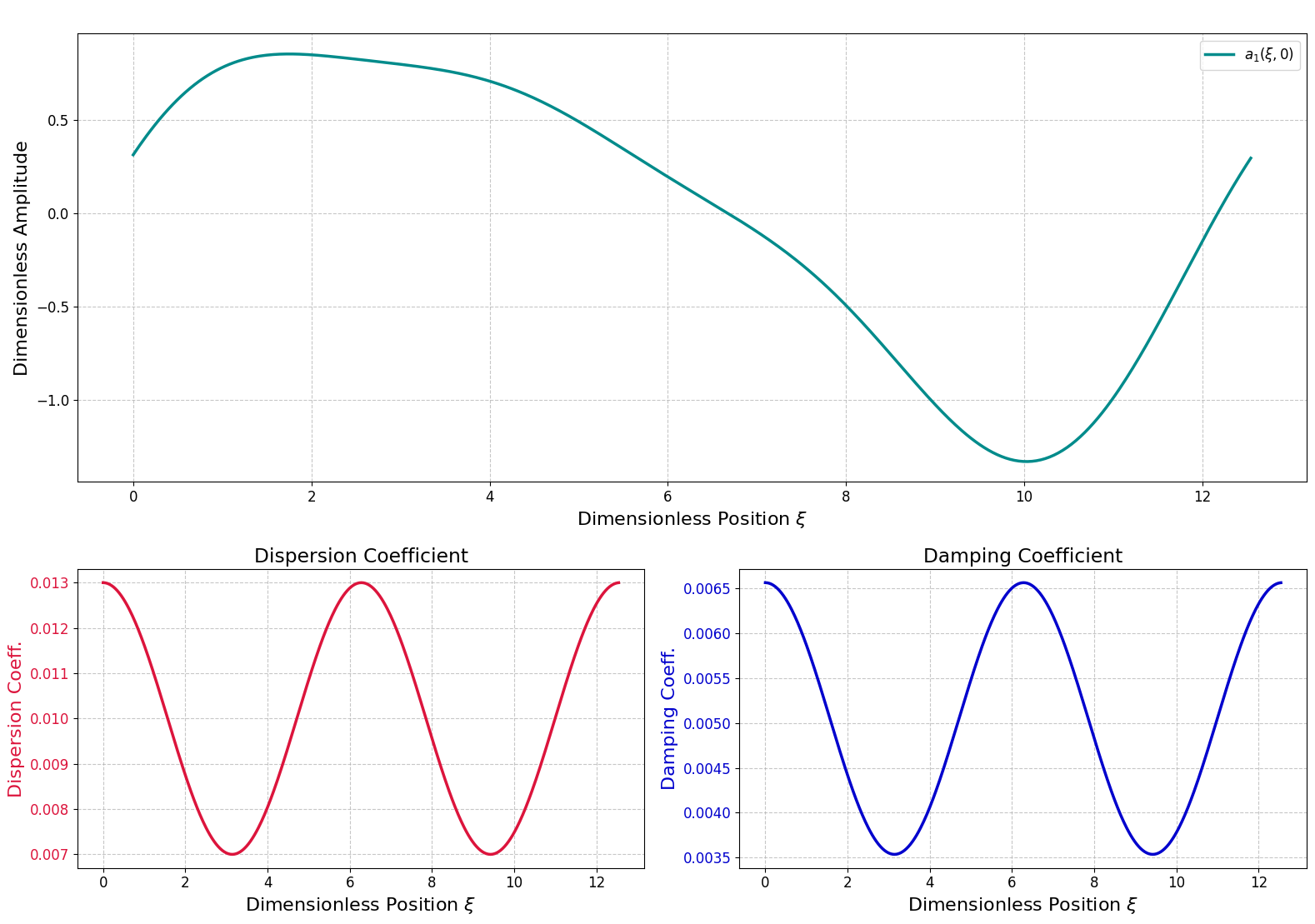}
\caption{Initial configuration for the representative case $\mathrm{Wo}=10$. 
Top: Multi-harmonic initial condition $a_1(\xi,0)$ composed of three Fourier modes. 
Bottom: Prescribed spatial profiles of the dispersion coefficient $\beta(\xi)$ (red) 
and damping coefficient $\eta(\xi)$ (blue), illustrating the geometric modulation 
introduced at the model level. The numerical evolution employs the spatial means 
$\bar{\beta}$ and $\bar{\eta}$ in the linear operator.}
\label{fig:initial_setup}
\end{figure}

The simulation was performed for a representative case with a Womersley number of $\mathrm{Wo}=10$. The initial conditions and spatially-modulated coefficients for this run are presented in Figure~\ref{fig:initial_setup}. The top panel shows the initial waveform, $a_1(\xi, 0)$, a composite of three harmonics resulting in an asymmetric profile with a dominant peak and a broader, shallower trough. The bottom panel displays the periodic functions for the dispersion coefficient, $\beta(\xi)$, and the damping coefficient, $\eta(\xi)$. BBoth coefficients vary in-phase with one another over two full geometric periods within the computational domain, illustrating the prescribed geometric modulation at the model level. 
In the numerical time integration, the spatial means $\bar{\beta}$ and $\bar{\eta}$ define the effective dispersive and dissipative parameters used in the linear propagator, so the plotted profiles represent the imposed geometric parameterization rather than a pointwise-varying operator during time stepping.

\begin{figure}[htbp]
\centering
\includegraphics[width=1\textwidth]{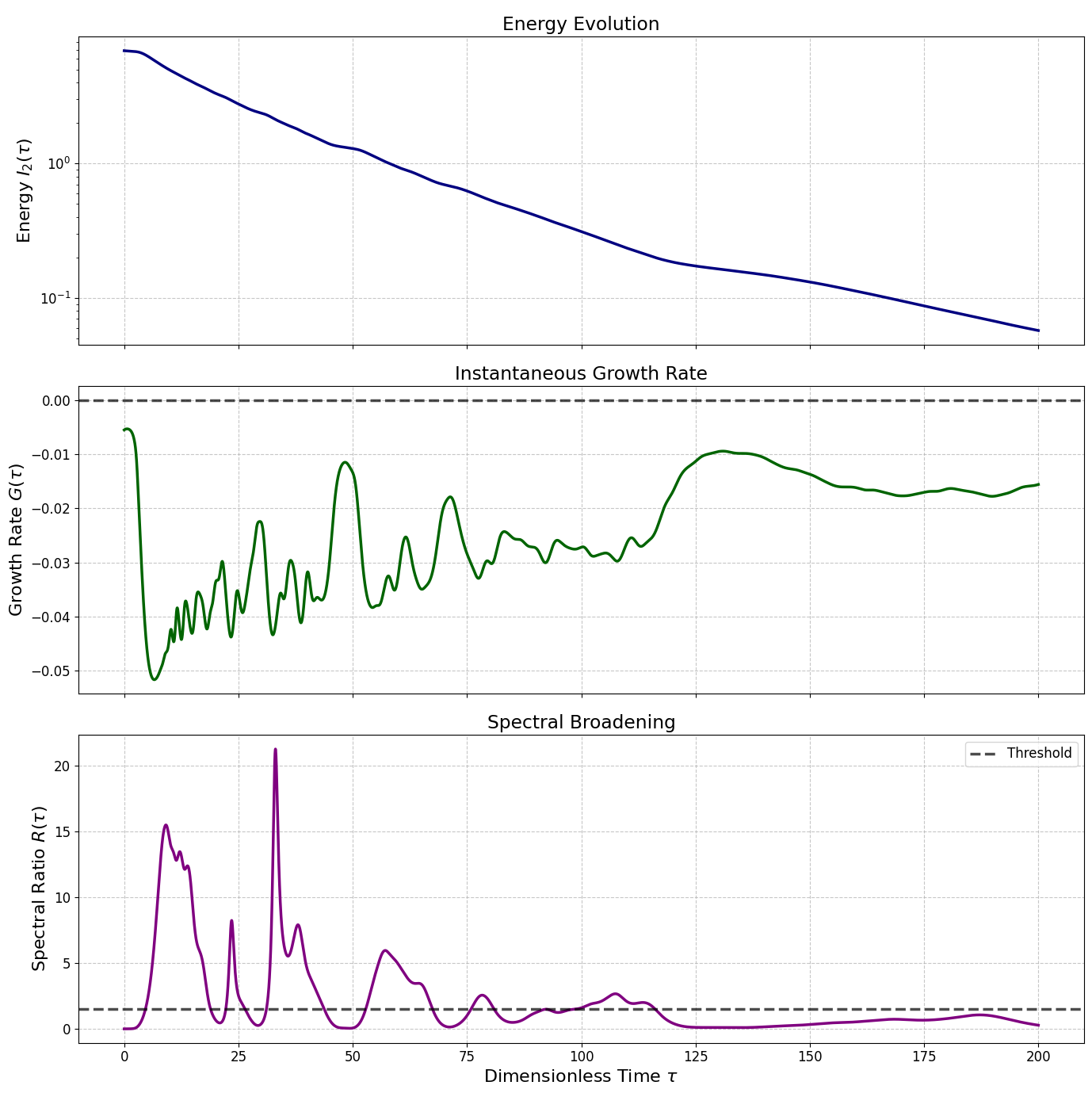}
\caption{Primary instability diagnostics for $\mathrm{Wo}=10$. 
Top: Total wave energy $I_2(\tau)$ (semi-log scale), showing monotonic decay. 
Middle: Instantaneous logarithmic growth rate $G(\tau)$, which remains negative 
for all $\tau$, confirming global stability. 
Bottom: Spectral broadening ratio $R(\tau)$, measuring redistribution of energy 
toward higher wavenumbers. A sharp transient peak precedes long-term damping, 
indicating efficient but non-exponential spectral transfer.}
\label{fig:diagnostics}
\end{figure}

The global dynamics of the system are summarized by the primary instability diagnostics in Figure~\ref{fig:diagnostics}. As shown in the top panel, the total wave energy, $I_2(\tau)$, decays by approximately two orders of magnitude over the simulation, indicating that the system is globally stable. The middle panel confirms this, showing that the instantaneous growth rate, $G(\tau)$, remains negative for all time; it undergoes large oscillations initially before settling to a quasi-steady negative value. In stark contrast, the bottom panel reveals a highly transient, nonlinear process: the spectral broadening ratio, $R(\tau)$, exhibits a sharp peak reaching a value greater than 20 within the first 40 time units. This demonstrates a rapid and significant transfer of energy to high wavenumbers that dramatically precedes the long-term energy decay.

\begin{figure}[htbp]
\centering
\includegraphics[width=1\textwidth]{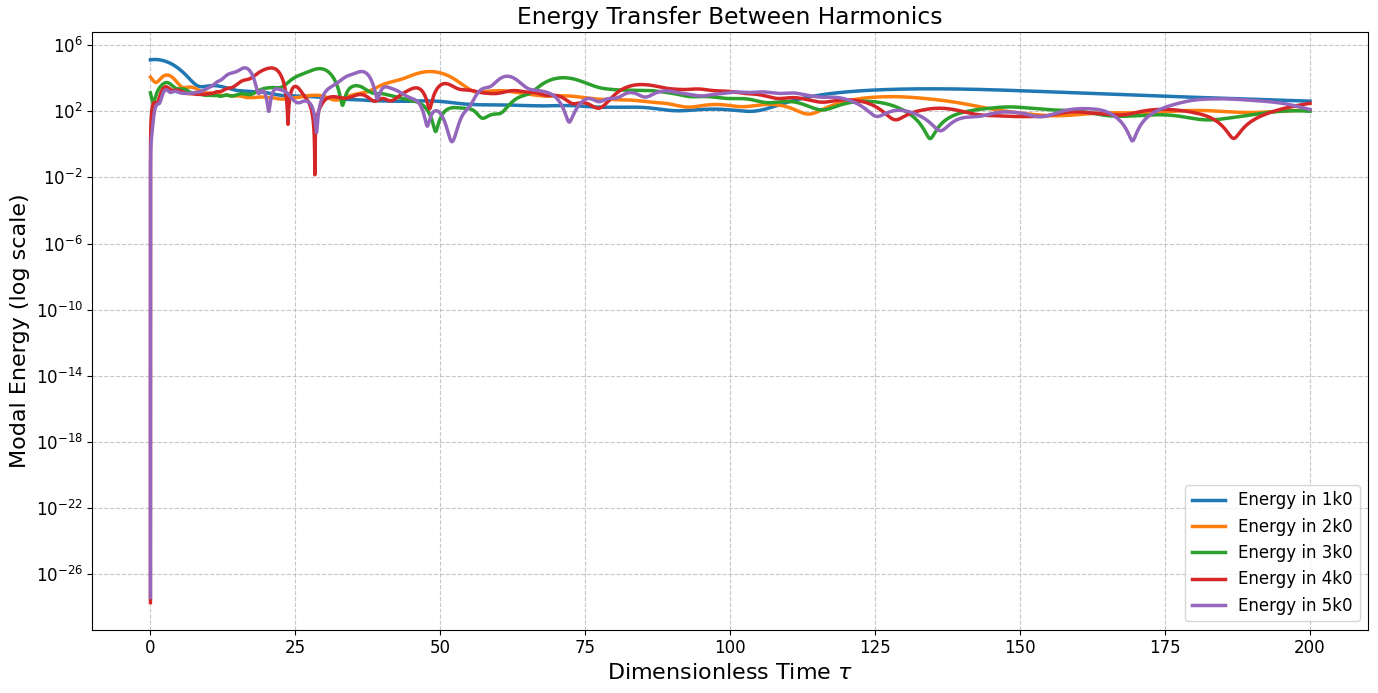}
\caption{
Temporal evolution of modal energies for the first five harmonics 
($k_0$ through $5k_0$) at $\mathrm{Wo}=10$. 
Rapid initial redistribution leads to near equipartition among low-order modes, 
followed by coupled decay.}
\label{fig:modal_energy}
\end{figure}

The mechanism of this energy transfer is detailed in Figure~\ref{fig:modal_energy}, which tracks the energy in the first five Fourier harmonics. The plot shows a rapid initial redistribution of energy, where all five modes achieve a state of near equipartition within the first few time units. Subsequently, the energy in all tracked modes undergoes a coupled, fluctuating decay. No single mode exhibits sustained exponential growth, which is consistent with the global stability of the system.

\begin{figure}[htbp]
\centering
\includegraphics[width=1\textwidth]{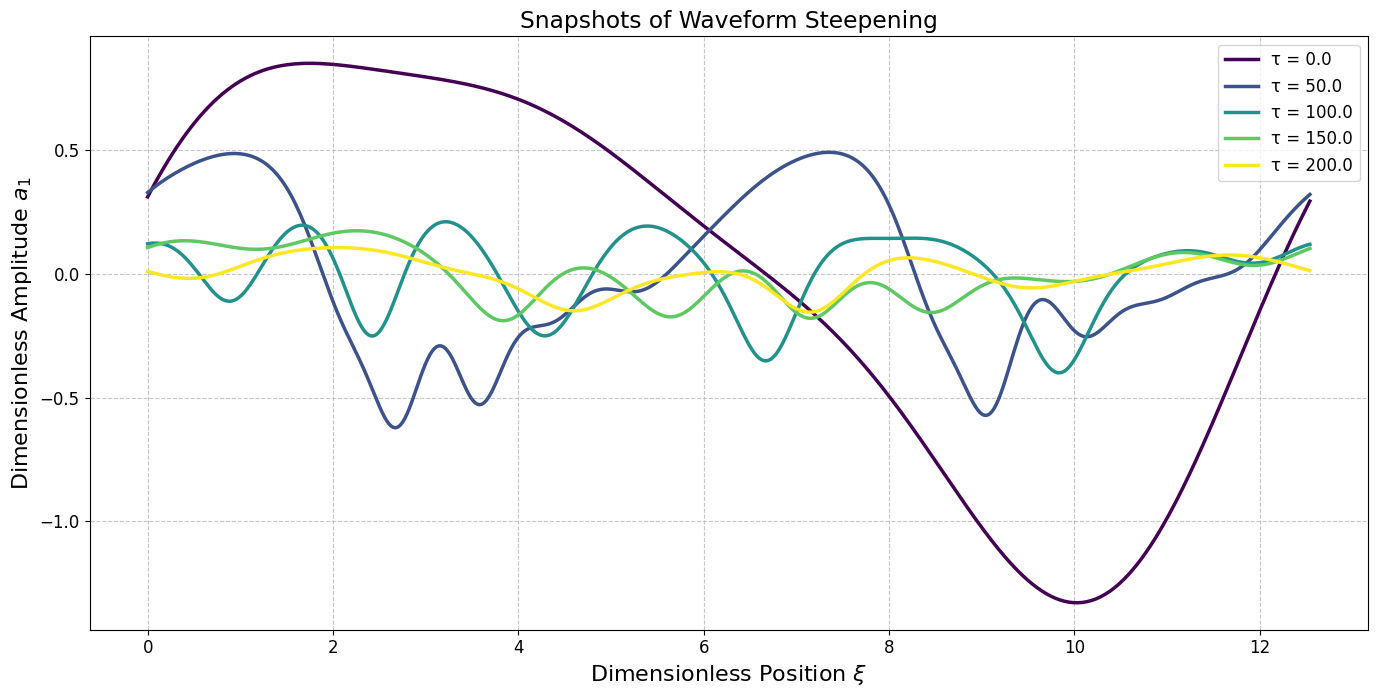}
\caption{Waveform evolution at selected times for $\mathrm{Wo}=10$. 
The initially smooth multi-harmonic profile develops short-wavelength oscillations 
during the transient redistribution phase, followed by gradual attenuation of both 
amplitude and fine-scale structure as dissipation dominates.}
\label{fig:wave_shape}
\end{figure}

The physical effect of the transient energy cascade on the waveform's shape is visualized in the snapshots of Figure~\ref{fig:wave_shape}. The initially smooth profile at $\tau=0.0$ undergoes a marked increase in short-wavelength oscillatory content seen at $\tau=50.0$, which is characterized by the emergence of numerous high-frequency oscillations and sharp gradients. The subsequent snapshots at $\tau=100.0$, $150.0$, and $200.0$ illustrate a process of gradual damping, where both the overall amplitude and the fine-scale structures are progressively attenuated.

\begin{figure}[htbp]
\centering
\includegraphics[width=1\textwidth]{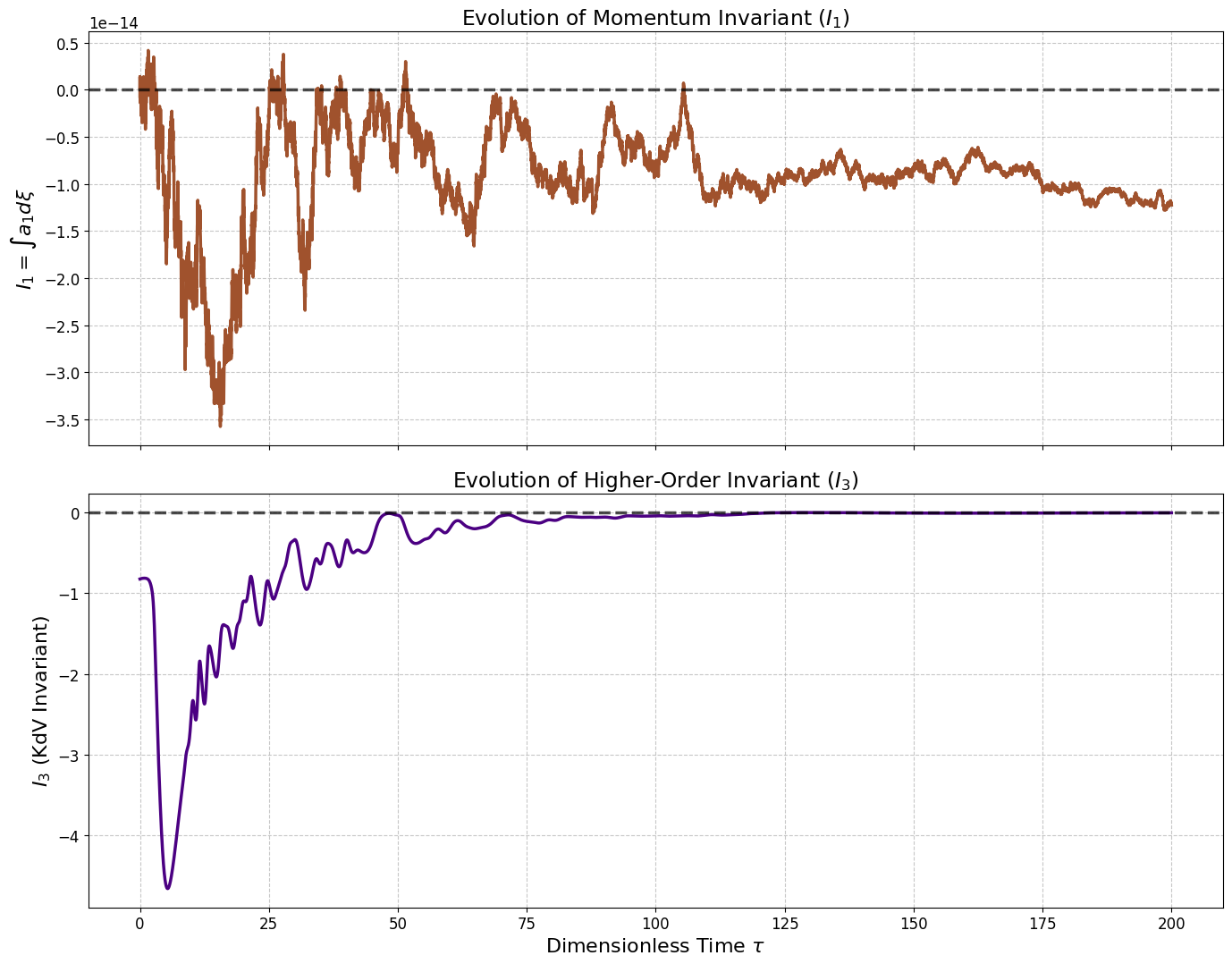}
\caption{Evolution of classical KdV invariants for $\mathrm{Wo}=10$. 
Top: Momentum-like invariant $I_1$. 
Bottom: Higher-order invariant $I_3$. 
Deviation from constant values confirms that the system is non-conservative due 
to dispersive and dissipative effects, in contrast to the integrable constant-coefficient 
KdV limit.}
\label{fig:invariants}
\end{figure}

Figure~\ref{fig:invariants} demonstrates the non-conservative nature of the system by tracking the evolution of the first and third classical KdV invariants. The momentum-like invariant, $I_1$ (top panel), deviates immediately from its initial zero value and undergoes sustained, decaying oscillations around a negative mean. The higher-order invariant, $I_3$ (bottom panel), experiences a sharp initial plunge before oscillating back towards a near-zero steady-state value. The dynamic behavior of both quantities confirms that the geometric terms are actively modulating the integral properties of the flow.

\begin{figure}[htbp]
\centering
\includegraphics[width=1\textwidth]{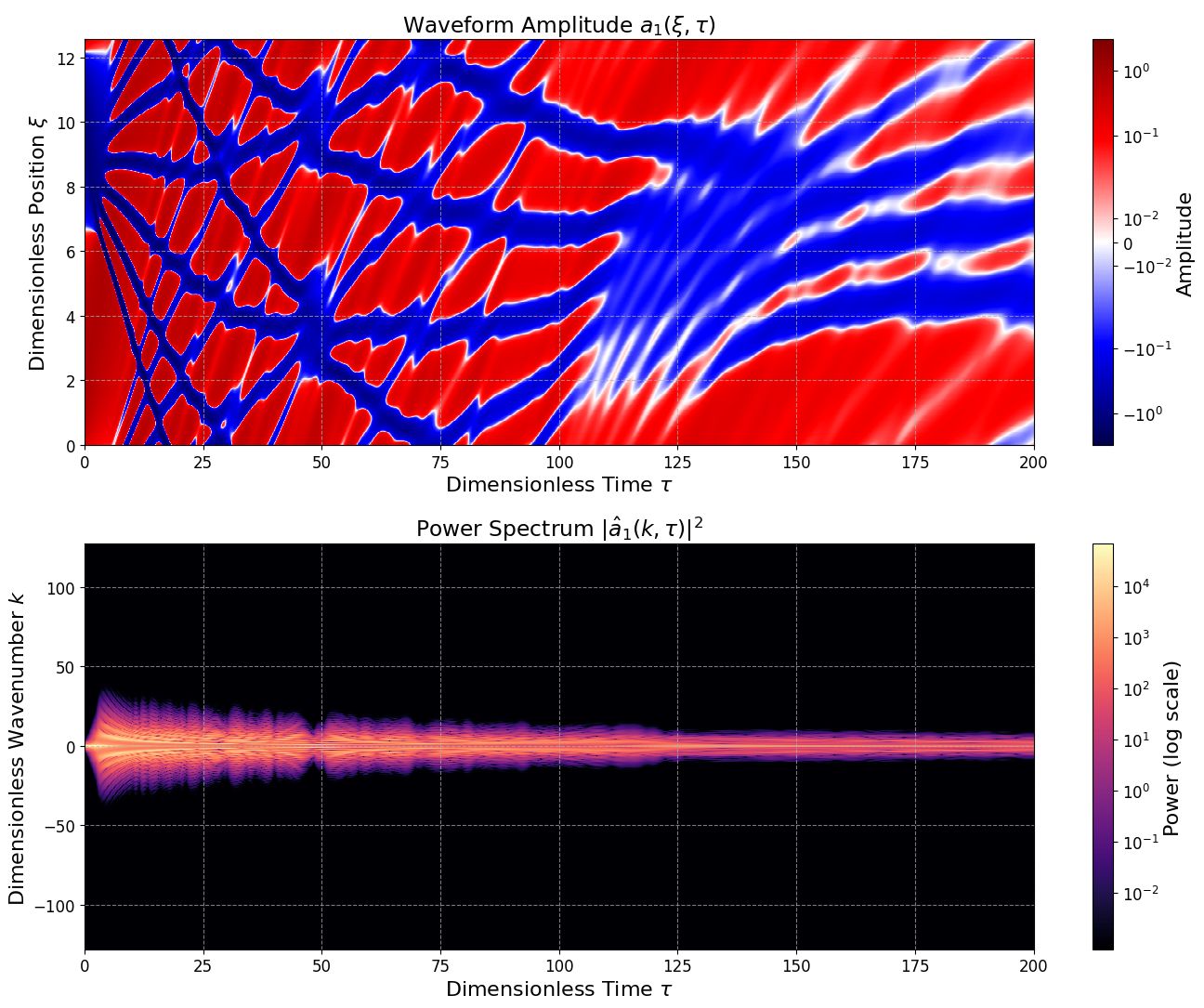}
\caption{Spatiotemporal evolution for $\mathrm{Wo}=10$. 
Top: Field $a_1(\xi,\tau)$ showing breakdown of large-scale coherence into 
small-scale structures during the transient phase. 
Bottom: Corresponding power spectrum $|\hat{a}_1(k,\tau)|^2$, demonstrating 
rapid broadening across wavenumbers before global decay reduces spectral intensity.}
\label{fig:evolution}
\end{figure}

A comprehensive spatiotemporal overview of the simulation is provided in Figure~\ref{fig:evolution}. The top panel, showing $a_1(\xi, \tau)$, reveals that the initial, large-scale coherent wave structures break down into a complex field of smaller, interacting wave packets. The bottom panel presents the corresponding power spectrum evolution, or spectrogram. This plot clearly visualizes the nearly instantaneous transfer of energy from the initial discrete modes to a broad, continuous spectrum of wavenumbers. This broadband state persists while its overall intensity gradually fades, visually confirming the global, system-wide damping over time.

\begin{figure}[htbp]
\centering
\includegraphics[width=1\textwidth]{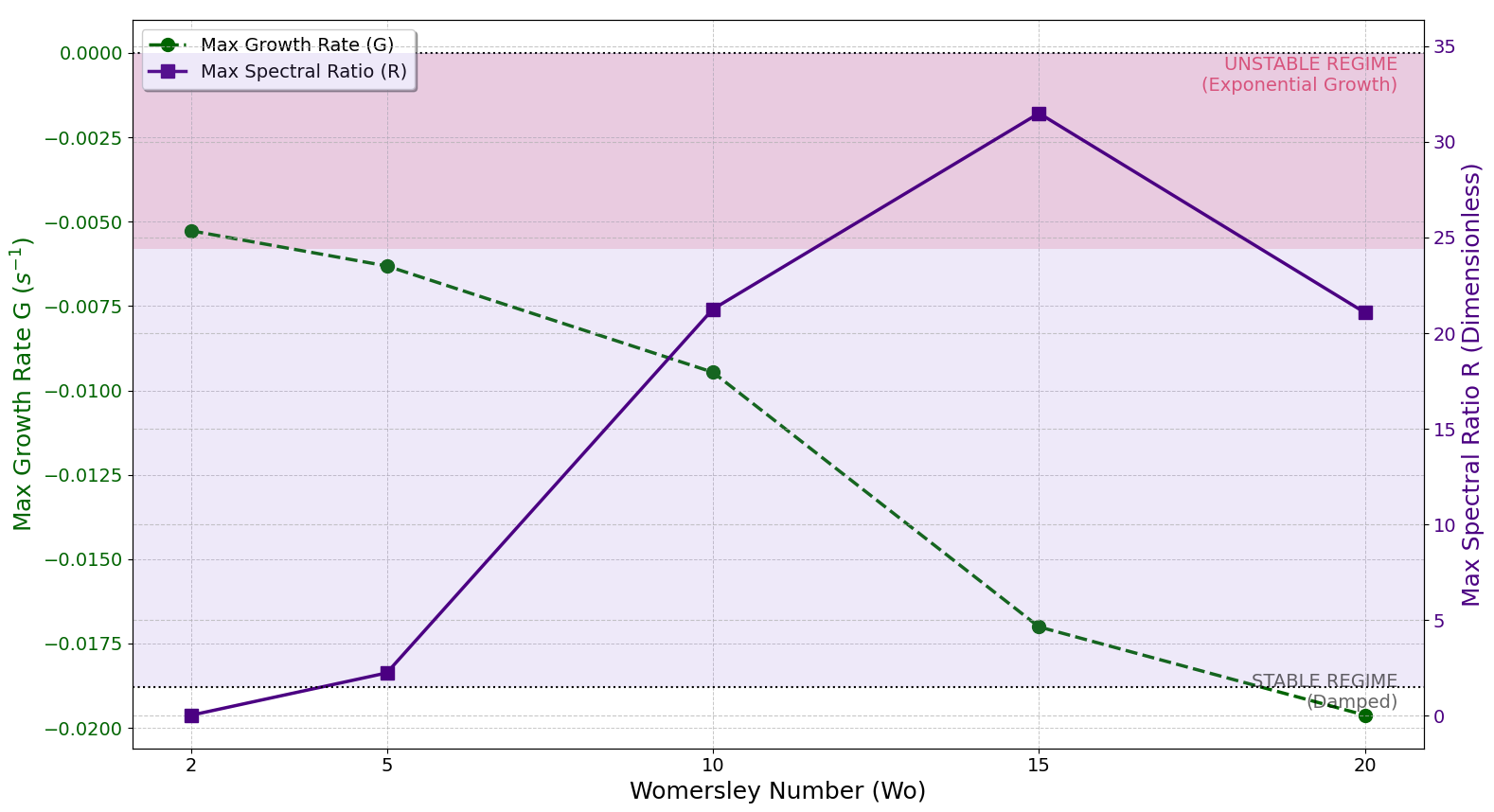}
\caption{Parameter sweep across $\mathrm{Wo}\in\{2,5,10,15,20\}$. 
Left axis: Maximum late-time growth rate $G$, which remains negative 
($G<0$) for all cases, indicating absence of exponential instability. 
Right axis: Maximum transient spectral broadening ratio $R$, showing a 
non-monotonic dependence on $\mathrm{Wo}$ with a peak near $\mathrm{Wo}=15$. 
Shaded regions indicate regimes of exponential growth ($G>0$, not observed) 
and broadband redistribution ($R>1.5$).}
\label{fig:sweep_results}
\end{figure}

The summary of the parameter sweep across the physiological range of Womersley numbers is presented in Figure~\ref{fig:sweep_results}. The plot synthesizes the primary instability metrics as a function of $\mathrm{Wo}$. The maximum late-stage growth rate, $G$, is plotted on the left vertical axis, while the maximum transient spectral broadening ratio, $R$, is plotted on the right. The results show that the growth rate $G$ remains negative for all tested Womersley numbers, indicating that the system is globally stable across this range. The magnitude of the decay increases with $\mathrm{Wo}$, suggesting stronger long-term damping at higher frequencies. In contrast, the spectral broadening ratio $R$ exhibits a non-monotonic dependence on the Womersley number. It rises from a low value at $\mathrm{Wo}=2$ to a distinct peak at $\mathrm{Wo}=15$, before decreasing again at $\mathrm{Wo}=20$. For all cases, the peak value of $R$ remains significantly above the broadband threshold of 1.5.

\begin{figure}[htbp]
\centering
\includegraphics[width=1.0\textwidth]{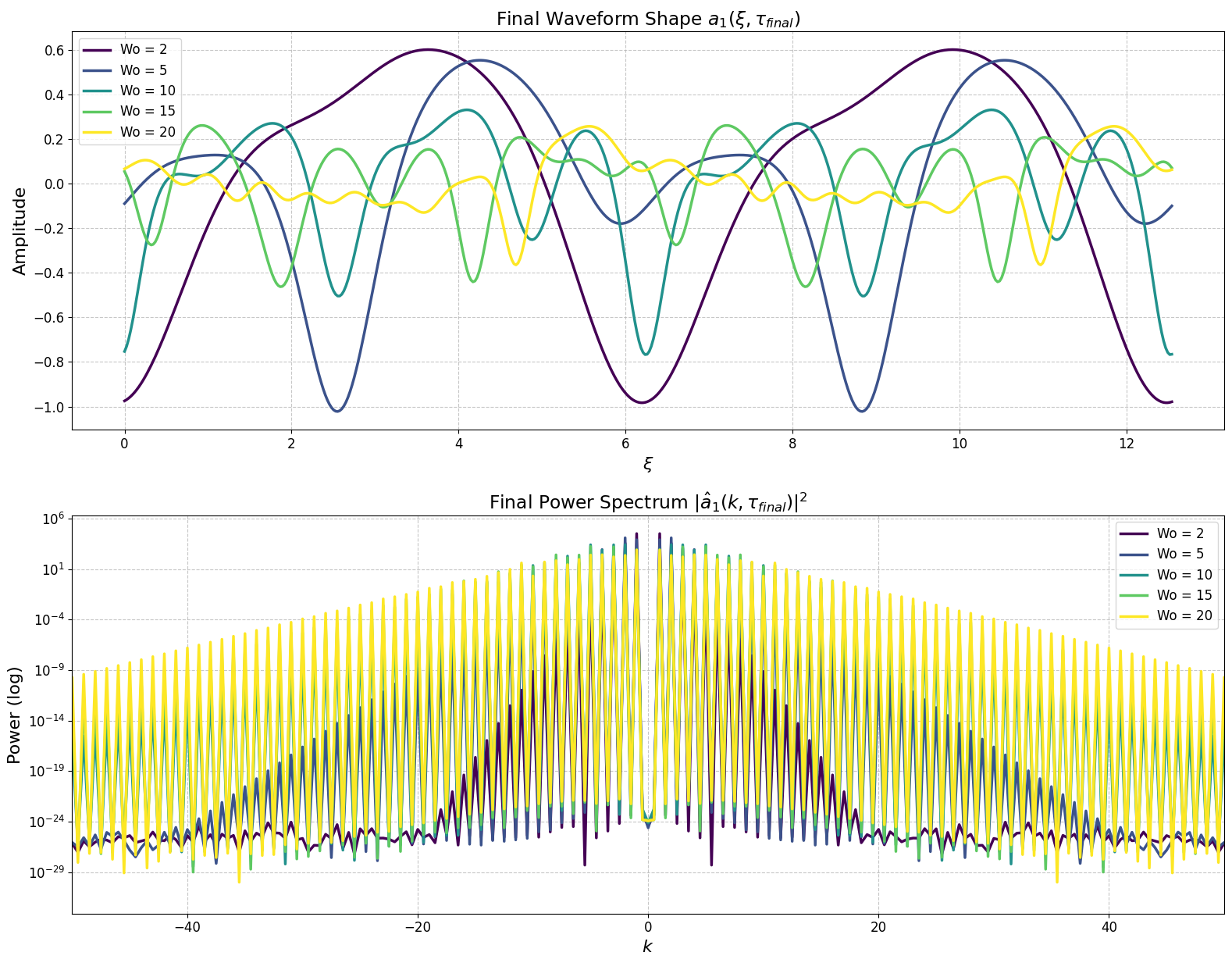}
\caption{Final-state comparison across Womersley numbers. 
Top: Final waveform $a_1(\xi,\tau_{\mathrm{final}})$ for each $\mathrm{Wo}$. 
Amplitude decreases at large $\mathrm{Wo}$ as damping strengthens. 
Bottom: Corresponding final power spectra $|\hat{a}_1(k,\tau_{\mathrm{final}})|^2$. 
Intermediate $\mathrm{Wo}$ values produce broader spectra, whereas low $\mathrm{Wo}$ 
cases retain energy concentrated at low wavenumbers.}
\label{fig:parametric_lines}
\end{figure}

\begin{figure}[htbp]
\centering
\includegraphics[width=1.0\textwidth]{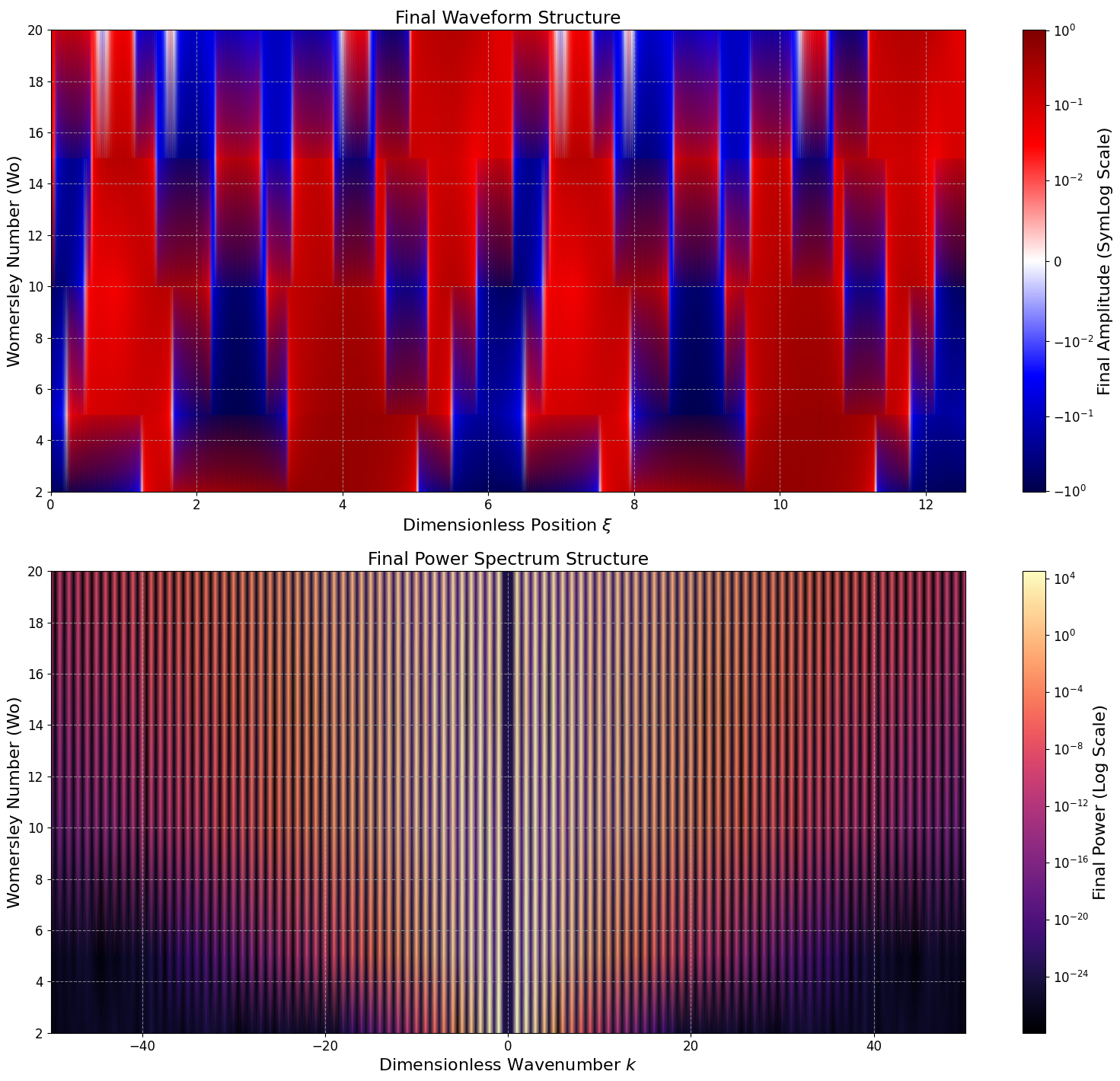}
\caption{Color-coded representation of final states across Womersley numbers. 
Top: Final waveform amplitude as a function of spatial coordinate $\xi$ and 
$\mathrm{Wo}$. 
Bottom: Final power spectrum magnitude $|\hat{a}_1(k,\tau_{\mathrm{final}})|^2$. 
Spectral width increases from $\mathrm{Wo}=2$ to approximately $\mathrm{Wo}=15$, 
then contracts at $\mathrm{Wo}=20$ due to stronger dissipation, consistent with 
the non-monotonic behavior of the spectral broadening ratio.}
\label{fig:parametric_structure}
\end{figure}

The parametric dependence of the system's final state on the Womersley number is detailed in Figures~\ref{fig:parametric_lines} and~\ref{fig:parametric_structure}. Figure~\ref{fig:parametric_lines} presents direct overlays of the final-state solutions. The top panel shows the final waveform, $a_1(\xi, \tau_{final})$, for each tested $\mathrm{Wo}$. The final amplitude is a non-monotonic function of $\mathrm{Wo}$, with the largest amplitude observed at $\mathrm{Wo}=2$ and the smallest at $\mathrm{Wo}=20$. The waveform complexity, characterized by the prevalence of short-wavelength oscillations, appears greatest for intermediate Womersley numbers. The bottom panel displays the corresponding final power spectra, $|\hat{a}_1(k, \tau_{final})|^2$. All cases result in a broadband spectrum, but the width and energy content at high wavenumbers are maximized for the $\mathrm{Wo}=15$ case, while the spectrum for $\mathrm{Wo}=2$ is the most narrowly concentrated at low wavenumbers.

A continuous visualization of these parametric trends is provided in Figure~\ref{fig:parametric_structure}. The top panel displays the final waveform structure as a heatmap, where the vertical axis represents the Womersley number. The transition from large-scale, high-amplitude coherent structures at low $\mathrm{Wo}$ to smaller, more complex, and lower-amplitude patterns at high $\mathrm{Wo}$ is evident. Similarly, the bottom panel shows the final power spectrum structure. This visualization clearly illustrates the progressive widening of the power spectrum as $\mathrm{Wo}$ increases from 2 to 15, followed by a global suppression of power at $\mathrm{Wo}=20$, confirming the resonant nature of the spectral broadening and the dominant effect of damping at high frequencies.

\section*{Discussion}

This study was designed to investigate the role of arterial geometry in the stability of pulsatile flow, specifically to determine if geometric features can actively drive instability. The comprehensive results from the numerical experiments, spanning a range of physiological Womersley numbers, provide a nuanced and physically rich answer. While the system does not exhibit exponential, runaway instability for the tested parameter regime, the findings conclusively establish that geometry functions as a powerful spectral catalyst. It actively and selectively generates flow complexity by governing a competition between parametric forcing, nonlinear dynamics, and frequency-dependent dissipation. A complete, evidence-based account of this mechanism can be constructed by addressing a sequence of logical inquiries that systematically connect the observed results to the underlying mathematical model.

First, what is the fundamental role of geometry within this theoretical framework? The system is modeled by the variable-coefficient fKdV equation, where the influence of geometry is explicitly encoded in the spatially-varying coefficients $\beta(\xi)$ and $\eta(\xi)$. A key finding, shown in Figure~\ref{fig:invariants}, is the non-conservation of the classical KdV invariants, $I_1$ and $I_3$. 
In the constant-coefficient KdV limit these quantities are strictly conserved, so their deviation here reflects the presence of dispersive and dissipative terms arising from geometry-parameterized coefficients. 
Although the numerical evolution employs the spatial means of $\beta(\xi)$ and $\eta(\xi)$ in the linear propagator, the resulting dynamics remain distinct from the integrable limit and therefore quantify the influence of geometry-induced parameter renormalization on spectral redistribution \cite{Bahloul2023, Azmi2025}.

\subsection*{Observability and experimental detectability}
The geometry-induced spectral redistribution discussed here is an \emph{axial-wavenumber} effect: energy is transferred from the initially imposed low-wavenumber content (set by $k_0$) toward higher axial wavenumbers, as quantified by the spectral broadening ratio $R(\tau)$ defined from the Fourier spectrum of $a_1(\xi,\tau)$. In the present nondimensionalization $\xi=x/L_0$, so an axial wavenumber $k$ corresponds to a dimensional wavelength $\lambda = 2\pi L_0/k$. Accordingly, the smallest axial length scale explicitly interrogated by the diagnostic (i.e., the ``high-$k$'' band above the cutoff $k_c$ used in defining $R$) is
\[
\lambda_{\min} \sim \frac{2\pi L_0}{k_c}.
\]
With the representative choices used herein ($k_c=O(1$--$10)$), $\lambda_{\min}$ is $O(L_0)$; for large conduit arteries a natural choice is $L_0\sim R_0$, implying $\lambda_{\min}=O(R_0)$, i.e.\ millimetric-to-centimetric axial scales for $R_0\sim 5$--$15$\,mm.

This scale separation indicates that the predicted redistribution is, in principle, accessible to current velocity-measurement modalities in large arteries. In vivo, 4D flow MRI provides time-resolved volumetric velocity fields and is suited to resolving axial variations at millimetric spatial resolution over the cardiac cycle; the model prediction would manifest as a transient increase in higher-$k$ content of the \emph{axial} velocity (or derived wall shear stress) along the vessel centerline or along a centerline-aligned curvilinear coordinate. High-frame-rate Doppler ultrasound, while spatially more limited, directly reports spectral broadening of the velocity signal and can therefore probe time-localized changes in spectral content over the cardiac period. Finally, in vitro particle image velocimetry (PIV) in compliant or rigid phantoms can provide the spatial and temporal resolution needed to resolve the full axial spectrum and to validate the predicted redistribution mechanism under controlled geometric modulation. The present results therefore support a measurable signature at conduit-artery scales, while acknowledging that resolving substantially smaller scales than $\lambda_{\min}$ would require dedicated experimental designs beyond standard clinical protocols.

Given that the system is actively forced but remains globally stable—with a negative energy growth rate $G$ across all tested Womersley numbers (Figure~\ref{fig:sweep_results})—what is the primary physical consequence of this forcing? The results demonstrate that the forcing drives a highly efficient, transient spectral cascade. This is observed directly as the rapid peak in the spectral broadening ratio $R$ (Figure~\ref{fig:diagnostics}) and is visualized as the breakdown of the initially smooth waveform (Figure~\ref{fig:initial_setup}) into a field of complex, high-frequency structures (Figure~\ref{fig:wave_shape} and~\ref{fig:evolution}). This process, whereby energy is rapidly redistributed from the initial low-wavenumber harmonics to a broad continuum of modes (Figure~\ref{fig:modal_energy}), is governed by the interaction between the nonlinear steepening term, $\alpha_0 a_1\partial_\xi a_1$, and the parametrically-forced dispersive term, $\beta_0\varepsilon_\beta\cos(q\xi)\partial^3_\xi a_1$. Unlike an integrable system which balances these effects to form stable solitary waves, the geometry-parameterized dispersive response disrupts this balance, preventing the formation of stable structures and instead promoting a chaotic-like state of high complexity. This provides a clear reduced-order mechanism for one class of geometry-sensitive unsteady behavior reported in other computational models \cite{Freidoonimehr2021, Yi2022}, and is consistent with the non-Kolmogorov spectral cascades and near-wall coherent structures observed in our prior high-fidelity simulations of patient-specific neurovascular geometries \cite{Saqr2022SR, Saqr2022AIP}.

This interpretation also aligns, at a broader phenomenological level, with prior CFD studies showing that vascular geometry can reorganize unsteady pulsatile flow in ways that are strongly phase- and morphology-dependent. In idealized curved or bifurcating configurations, geometric variation has been shown to modify flow partitioning, alter oscillatory wall-shear patterns, shift stagnation dynamics, and break planar symmetry over the cardiac cycle \cite{Pivkin2005,Papaharilaou2002,Sherwin2000,He1993}. In more irregular or diseased geometries, numerical studies have further reported enhanced disturbance levels, transitional behavior, broadband velocity fluctuations, and Doppler-observable spectral broadening under pulsatile conditions \cite{Lee2007,Lee2008,Wong2008,Wong2009,Varghese2007,Khoshniat2005}. These results support the general premise that arterial geometry influences the temporal organization and frequency content of pulsatile flow. Nevertheless, those studies do not isolate the specific mechanism identified here. The role of the present model is therefore explanatory. This study extracts and dissects an interpretable mechanism by which geometry can act as a spectral modulator in the main human arteries.

\subsection*{Harmonic Regulation}
How does the Womersley number, a parameter of the underlying base flow, regulate this geometrically-driven process? The parameter sweep results in Figure~\ref{fig:sweep_results} reveal a distinct dual role for $\mathrm{Wo}$. On one hand, it governs the overall rate of dissipation, as seen by the increasingly negative growth rate $G$ at higher $\mathrm{Wo}$. On the other hand, it critically tunes the efficiency of the spectral cascade. The non-monotonic, resonant behavior of the spectral ratio $R$, which peaks at an intermediate value of $\mathrm{Wo}=15$, is the key evidence for this tuning. The mathematical basis for this resonance is the scaling of the baseline dispersion coefficient, $\beta_0 \propto \mathrm{Wo}^{-2}$, as defined in the `Methods`. As $\mathrm{Wo}$ increases, the intrinsic dispersion of the wave weakens, making it more susceptible to nonlinear steepening and parametric forcing. The peak at $\mathrm{Wo}=15$ thus represents an optimal matching between the wave's intrinsic timescale, set by $\mathrm{Wo}$, and the fixed spatial period of the geometric forcing, set by the wavenumber $q$. Conceptually, this behavior is analogous to the resonant "tongues" predicted by Floquet stability theory for parametrically forced systems \cite{Kern2024}, where instability is maximized at specific forcing frequencies.

This leads to a non-intuitive but critical observation when examining the final, asymptotic states of the system. How can the most spectrally energetic and complex state at $\mathrm{Wo}=15$ result in a physical waveform with a lower peak amplitude than the simpler state at $\mathrm{Wo}=2$? The parametric structure plots in Figure~\ref{fig:parametric_structure} provide the definitive answer. The apparent paradox is resolved by distinguishing between energy concentration and energy distribution. At low $\mathrm{Wo}$, the spectral cascade is inefficient; energy remains concentrated in the first few low-wavenumber modes, manifesting as simple, large-amplitude coherent waves. At the resonant peak of $\mathrm{Wo}=15$, the cascade is maximally efficient, shattering the initial wave and distributing its energy across a vast range of high-frequency modes. While the total spectral energy is high, its distribution among many small-scale, out-of-phase structures results in a lower peak physical amplitude, analogous to how broadband white noise has a lower peak pressure than a powerful, low-frequency pure tone of the same total acoustic power.

Finally, what happens at very high Womersley numbers, causing the trend to reverse after the resonant peak? The answer lies in the second role of $\mathrm{Wo}$: controlling damping. As seen across all parametric plots (Figures~\ref{fig:sweep_results}, \ref{fig:parametric_lines}, and \ref{fig:parametric_structure}), the solution at $\mathrm{Wo}=20$ is significantly attenuated. At these high frequencies, the damping term $\eta(\xi)(-\partial_\xi^2)^{\frac{1+\gamma_0}{2}}a_1$ becomes highly effective at dissipating the very small-scale structures that the resonant cascade efficiently creates. The final state of the system is therefore a snapshot of the dynamic equilibrium reached between the rate of complexity generation, tuned by the dispersive properties, and the rate of complexity dissipation, governed by the damping properties.

It is important to acknowledge the limitations of the present model. By design, our framework isolates geometry as the sole parametric forcing agent and assumes a rigid wall. 

The effect of wall compliance may be estimated dimensionally by comparing the characteristic pulsatile flow velocity $U$ to the characteristic pulse-wave speed associated with pressure--area coupling. A standard scaling for the latter is the Moens--Korteweg wave speed,$
c_w \sim \sqrt{\frac{E h}{\rho R_0}}$, where $E$ is the effective Young's modulus of the arterial wall, $h$ its thickness, $\rho$ the fluid density, and $R_0$ the reference radius. This introduces a nondimensional compliance parameter $ \chi \equiv \frac{U}{c_w}$, which measures the strength of wall-induced wave dynamics relative to advective and inertial transport. The reduced evolution equation used here corresponds to the rigid-wall limit of the derivation in the Supplementary Information; incorporation of fluid--structure interaction would modify the dispersive balance but not the leading-order asymptotic structure outlined in Supplementary Information.

For large conduit arteries (e.g., aorta, renal, carotid), physiological parameters satisfy $U \ll c_w$, implying $\chi \ll 1$. In this strongly sub-wave-speed regime, wall motion modifies the axial momentum balance primarily through $O(\chi)$ corrections to phase speed and through additional attenuation associated with pressure--area energy exchange. Importantly, the leading-order inertial--viscous balance that generates Womersley dispersion remains dominant. Consequently, compliance acts predominantly as a renormalization of the effective dispersive and dissipative response, shifting quantitative peak timing and amplitude, rather than eliminating the geometry-modulated phase-redistribution mechanism identified here. The author's prior experimental work has shown that wall compliance can have a significant, often attenuating, effect on the kinetic energy cascade and near-wall velocity oscillations in aneurysm models \cite{Tupin2020EF, Yamaguchi2022JAP}. Therefore, a complete picture of hemodynamic stability requires understanding the interplay between the geometry-driven spectral cascade identified here and the dissipative or modulating effects of wall elasticity. Future work should aim to incorporate wall compliance into this reduced-order framework, potentially as a time-dependent or nonlocal damping term, to explore this competitive dynamic.

\section*{Conclusion}

When geometry is parameterized through dispersive and dissipative coefficients within a reduced-order pulsatile-flow model, it produces a resonant, frequency-dependent redistribution of spectral energy. For the studied parameter range, this mechanism does not generate exponential instability ($G<0$ for all cases), but it significantly modifies the distribution of energy across axial wavenumbers. The efficiency of this redistribution exhibits a non-monotonic dependence on the Womersley number, with a distinct peak at intermediate values.

The Womersley number therefore plays a dual regulatory role: it governs both the intrinsic dispersive scaling and the strength of dissipation, determining whether energy remains concentrated at low wavenumber or is transiently redistributed toward smaller axial scales. While the present one-dimensional formulation does not resolve full three-dimensional hemodynamics or fluid–structure interaction, it isolates a fundamental mechanism by which geometry-dependent parameter variations can reshape spectral structure without inducing runaway instability.

These results provide a quantitative framework for interpreting how geometric complexity may influence spectral organization in pulsatile internal flows and suggest experimentally accessible diagnostics based on axial spectral redistribution.

\section*{Supplementary Material}
Supplementary Information provides a detailed derivation of the variable-coefficient fractional Korteweg--de Vries model used in this manuscript. Starting from the incompressible Navier--Stokes equations under axisymmetric long-wave assumptions, it documents the perturbation expansion about Womersley base flow, radial-mode projection, fractional damping formulation, nondimensionalization, geometric modulation and Womersley scaling of coefficients, limiting cases, and the homogenized effective-medium form implemented by the numerical solver.

\section*{Acknowledgement}
The author acknowledges the anonymous reviewer \#3 of this article for their insightful remarks during the peer-review process. 

\section*{Code Availability}
The simulation code, including scripts for numerical integration and data analysis, is publicly available on Google Colab at the following URL: \url{https://colab.research.google.com/drive/1j_mJpn_j4BUD4DVZNpbGee1YAF2bKS_y}. All data generated during the parametric sweep, along with the detailed simulation results for individual Womersley numbers, are archived in the corresponding repository and can be accessed via the linked notebooks.

\singlespacing

\section*{Nomenclature}
\refstepcounter{table}
\label{tab:nomenclature}
\noindent\textbf{Table \thetable. Nomenclature}

\vspace{0.5em}
\noindent\textbf{Physical Variables and Parameters}
\begin{center}
\begin{tabular}{@{}llp{8.5cm}@{}}
\toprule
\textbf{Symbol} & \textbf{Units} & \textbf{Description} \\
\midrule
\(A(x,t)\)      & Varies       & Physical wave amplitude (e.g., of cross-sectional area or flow rate). \\
\(\mathbf{v}\)  & m/s          & Fluid velocity vector. \\
\(v_z, v_r\)    & m/s          & Axial and radial velocity components in cylindrical coordinates. \\
\(p\)           & Pa           & Pressure. \\
\(t\)           & s            & Time. \\
\(x, z\)        & m            & Axial spatial coordinate. \\
\(\rho\)        & kg/m\(^3\)  & Fluid density. \\
\(\mu\)         & Pa\(\cdot\)s & Dynamic viscosity. \\
\(\nu\)         & m\(^2\)/s   & Kinematic viscosity (\(\nu = \mu/\rho\)). \\
\(R_0\)         & m            & Characteristic vessel radius. \\
\(\omega\)      & rad/s        & Angular frequency of pulsation. \\
\(C(x), D(x), K(x)\) & Varies & Spatially-varying physical coefficients for nonlinearity, dispersion, and fractional effects. \\
\(\gamma_0\)   & --           & Constant order of the Riesz fractional operator. \\
\bottomrule
\end{tabular}
\end{center}

\vspace{0.5em}
\noindent\textbf{Dimensionless Variables and Parameters}
\begin{center}
\begin{tabular}{@{}llp{8.5cm}@{}}
\toprule
\textbf{Symbol} & \textbf{Units} & \textbf{Description} \\
\midrule
\(a_1(\xi,\tau)\)& --           & Dimensionless wave amplitude. \\
\(\xi\)         & --           & Dimensionless axial coordinate. \\
\(\tau\)        & --           & Dimensionless time. \\
\(\mathrm{Wo}\) & --           & Womersley number, ratio of unsteady inertial to viscous forces. \\
\(A_0, L_0, T_0\)& Varies, m, s & Reference scales for amplitude, length, and time used in nondimensionalization. \\
\(\alpha(\xi)\) & --           & Dimensionless, spatially-varying coefficient for nonlinearity. \\
\(\beta(\xi)\)  & --           & Dimensionless, spatially-varying coefficient for dispersion. \\
\(\eta(\xi)\)   & --           & Dimensionless, spatially-varying coefficient for the fractional term. \\
\(\alpha_0, \beta_0, \eta_0, \gamma_0\) & -- & Baseline (constant) values of the dimensionless coefficients. \\
\(\varepsilon_i\) & --         & Modulation amplitudes for the geometric coefficients (\(i \in \{\alpha, \beta, \eta\}\)). \\
\(q\)           & --           & Dimensionless geometric wavenumber. \\
\(L_g\)         & --           & Dimensionless length of the computational domain. \\
\(k_0\)         & --           & Fundamental dimensionless wavenumber for the initial condition. \\
\bottomrule
\end{tabular}
\end{center}

\vspace{0.5em}
\noindent\textbf{Mathematical and Numerical Symbols}
\begin{center}
\begin{tabular}{@{}llp{8.5cm}@{}}
\toprule
\textbf{Symbol} & \textbf{Units} & \textbf{Description} \\
\midrule
\(\mathcal{D}^{\gamma_0}\) & -- & Riesz fractional derivative operator of constant order \(\gamma_0\). \\
\(\mathcal{L}, \mathcal{N}\) & -- & Continuous linear and nonlinear operators. \\
\(\mathbf{L}\)            & -- & Discretized linear operator (matrix form). \\
\(\hat{a}_1(k)\)          & -- & Fourier transform of the dimensionless amplitude \(a_1(\xi)\). \\
\(k\)                     & -- & Dimensionless wavenumber in Fourier space. \\
\(\Delta\tau, \Delta\xi\) & -- & Time step and spatial grid spacing for numerical simulation. \\
\(N\)                     & -- & Number of spatial collocation points. \\
\(I_1, I_2, I_3\)         & -- & Conserved quantities (invariants) of the classical KdV equation. \\
\(G\)                     & -- & Instantaneous growth rate of the wave energy (\(I_2\)). \\
\(R\)                     & -- & Spectral broadening ratio (high-frequency to low-frequency energy). \\
\bottomrule
\end{tabular}
\end{center}

\newpage
\appendix
\fancypagestyle{supplementaryrunning}{%
  \fancyhf{}
  \fancyhead[C]{\footnotesize\color{gray}\ifodd\value{page}K.M. Saqr \textbar\ Resonant spectral cascade in Womersley flow triggered by arterial geometry\else Accepted for Publication in Physics of Fluids\fi}
  \fancyfoot[C]{\thepage}
}
\pagestyle{supplementaryrunning}
\thispagestyle{supplementaryrunning}

\begin{center}
{\LARGE\textbf{Supplementary Information}}\\[0.35em]
{\large\textbf{Detailed Derivation of the Fractional KdV Model}}
\end{center}
\vspace{0.8em}

\renewcommand{\theequation}{A.\arabic{equation}}
\renewcommand{\theHequation}{A.\arabic{equation}}
\setcounter{equation}{0}

This appendix provides a graduate-level derivation of the one-dimensional variable-coefficient fractional Korteweg--de Vries (fKdV) equation used in the main text. The derivation proceeds from the incompressible Navier--Stokes equations in cylindrical coordinates, through a weakly nonlinear perturbation analysis, radial-mode projection, and non-dimensionalization, yielding a reduced-order model that incorporates geometric dispersion, nonlinear steepening, and fractional geometric damping.

\subsection*{A.1. Governing equations and geometric assumptions}

Blood is modeled as an incompressible Newtonian fluid with density $\rho$ and dynamic viscosity $\mu$. In cylindrical coordinates $(r,\theta,z)$, the velocity field is $\mathbf{v}=(v_r,v_\theta,v_z)$ and the pressure is $p$. The continuity and axial momentum equations are
\begin{align}
\frac{1}{r}\frac{\partial}{\partial r}(r v_r) + \frac{1}{r}\frac{\partial v_\theta}{\partial \theta}
+ \frac{\partial v_z}{\partial z} &= 0, \\
\rho\!\left(
\frac{\partial v_z}{\partial t}
+ v_r \frac{\partial v_z}{\partial r}
+ \frac{v_\theta}{r}\frac{\partial v_z}{\partial \theta}
+ v_z \frac{\partial v_z}{\partial z}
\right)
&= -\frac{\partial p}{\partial z}
+ \mu\!\left[
\frac{1}{r}\frac{\partial}{\partial r}\!\left(r\frac{\partial v_z}{\partial r}\right)
+ \frac{1}{r^2}\frac{\partial^2 v_z}{\partial \theta^2}
+ \frac{\partial^2 v_z}{\partial z^2}
\right]. 
\end{align}

We adopt the following assumptions, standard in pulsatile hemodynamics and long-wave asymptotics:
\begin{itemize}
    \item \textbf{Axisymmetry}: $\partial/\partial\theta = 0$ and $v_\theta = 0$;
    \item \textbf{Slenderness}: $R_0/L_0 = \delta \ll 1$ for characteristic radius $R_0$ and axial scale $L_0$;
    \item \textbf{Mild curvature/torsion}: curvature $\kappa(x)$ and torsion $\tau(x)$ satisfy $\kappa R_0,\tau R_0 \ll 1$ and act through slow axial modulation of coefficients.
\end{itemize}

Under axisymmetry, (A.2) simplifies to
\begin{equation}
\rho\!\left(
\frac{\partial v_z}{\partial t}
+ v_r\frac{\partial v_z}{\partial r}
+ v_z\frac{\partial v_z}{\partial z}
\right)
= -\frac{\partial p}{\partial z}
+ \mu\!\left[
\frac{1}{r}\frac{\partial}{\partial r}\!\left(r\frac{\partial v_z}{\partial r}\right)
+ \frac{\partial^2 v_z}{\partial z^2}
\right].
\end{equation}

\subsection*{A.2. Base oscillatory Womersley flow}

The zeroth-order solution is the classical Womersley flow driven by a harmonic pressure gradient
\begin{equation}
\frac{\partial p_0}{\partial z} = \Re \left\{ \hat{P}(x)e^{i\omega t} \right\}.
\end{equation}
Under Womersley conditions, the convective terms vanish identically:
\begin{equation}
\rho \frac{\partial v_0}{\partial t}
=
-\frac{\partial p_0}{\partial z}
+ \mu \left[
\frac{1}{r}\frac{\partial}{\partial r}\!\left(r\frac{\partial v_0}{\partial r}\right)
\right].
\end{equation}
Its solution is the known Womersley profile
\begin{equation}
v_0(r,t;x)
=
\Re\!\left[
\frac{\hat{P}(x)}{i\rho\omega}
\left(
1 - 
\frac{J_0(i^{3/2}\mathrm{Wo}\, r/R_0)}
     {J_0(i^{3/2}\mathrm{Wo})}
\right)
e^{i\omega t}
\right],
\end{equation}
where $\mathrm{Wo}=R_0\sqrt{\omega/\nu}$ is the Womersley number, $\nu=\mu/\rho$.

We seek an evolution equation for perturbations $v'$, writing
\begin{equation}
v_z(r,x,t) = v_0(r,t;x) + v'(r,x,t).
\end{equation}

\subsection*{A.3. Perturbation ansatz and long-wave scaling}

We introduce a small parameter $\varepsilon\ll 1$ and expand the perturbation in the leading radial eigenmode $\phi(r)$:
\begin{equation}
v'(r,x,t) = \varepsilon A(x,t)\phi(r) + O(\varepsilon^2).
\end{equation}

We assume Womersley oscillations act on a fast timescale $t_0=t$ while the perturbation evolves on a slow timescale $t_1=\varepsilon t$. Using long-wave scaling,
\[
\frac{\partial}{\partial x}\sim O(\varepsilon),\quad 
\frac{\partial}{\partial t_1}\sim O(\varepsilon),
\]
we let derivatives transform as
\begin{equation}
\partial_t \to \partial_{t_0} + \varepsilon\partial_{t_1}.
\end{equation}

\subsection*{A.4. Radial eigenproblem and projection}

Expanding the pressure as $p=p_0+\varepsilon p_1+\cdots$, the $O(\varepsilon)$ perturbation satisfies
\begin{equation}
\rho\phi\,\frac{\partial A}{\partial t_0}
=
-\frac{\partial p_1}{\partial z}
+ \mu A \mathcal{L}_r[\phi],
\qquad
\mathcal{L}_r[\phi] = \frac{1}{r}\frac{d}{dr}\!\left(r\frac{d\phi}{dr}\right).
\end{equation}

We introduce the cross-sectional inner product
\begin{equation}
\langle f,g\rangle = \int_0^{R_0} f(r)g(r)\, r\, dr,
\end{equation}
and normalize $\phi$ such that $\langle \phi,\phi\rangle=1$.

Projecting the full $O(\varepsilon^2)$ equation onto $\phi(r)$ produces a solvability condition yielding a scalar amplitude equation of the form
\begin{equation}
\frac{\partial A}{\partial t}
+ C(x)A\frac{\partial A}{\partial x}
+ D(x)\frac{\partial^3 A}{\partial x^3}
+ K(x)\mathcal{D}^{\gamma(x)}A
=0,
\end{equation}
where $C(x)$, $D(x)$, and $K(x)$ are determined by radial integrals involving $\phi$ and geometric corrections.

\subsection*{A.5. Nonlinear term}

The leading nonlinearity arises from
\[
v'\frac{\partial v'}{\partial z}
= \varepsilon^2 A\,\frac{\partial A}{\partial x}\phi^2(r) + \cdots
\]
and additional contributions involving $v_r\partial_r v'$.
After projection, this yields a term
\begin{equation}
C(x)A\frac{\partial A}{\partial x},
\end{equation}
with
\[
C(x)\sim \rho\int_0^{R_0}\!\phi^2(r)\phi(r)\, r\, dr,
\]
up to geometry-induced corrections.

\subsection*{A.6. Dispersive term}

Long-wave axial variations and geometry introduce higher-order axial derivatives. The leading dispersive correction takes the form
\begin{equation}
D(x)\frac{\partial^3 A}{\partial x^3},
\end{equation}
where $D(x)$ depends on viscosity, curvature, torsion, and Womersley-dependent radial coupling.

\subsection*{A.7. Fractional geometric damping}

Geometric roughness, tortuosity, and distributed compliance introduce multiscale nonlocal damping. This is modeled by a Riesz fractional operator
\begin{equation}
\mathcal{F}\{\mathcal{D}^{\gamma}A\}(k) = |k|^{1+\gamma}\hat{A}(k),
\end{equation}
giving the term
\begin{equation}
K(x)\mathcal{D}^{\gamma(x)}A.
\end{equation}

In this work, we set $\gamma(x)=\gamma_0=0$ to isolate geometric forcing through $D(x)$ and $K(x)$.

\subsection*{A.8. Non-dimensionalization}

Introduce
\[
\xi = \frac{x}{L_0},\qquad \tau = \omega t,\qquad A(x,t)=A_0a_1(\xi,\tau),
\]
so that
\[
\partial_t = \omega\partial_\tau,\quad
\partial_x = \frac{1}{L_0}\partial_\xi,\quad
\partial_x^3 = \frac{1}{L_0^3}\partial_\xi^3.
\]

Equation (A.12) becomes
\begin{equation}
\frac{\partial a_1}{\partial \tau}
+ \alpha(\xi)a_1\frac{\partial a_1}{\partial \xi}
+ \beta(\xi)\frac{\partial^3 a_1}{\partial \xi^3}
+ \eta(\xi)(-\partial_\xi^2)^{\frac{1+\gamma_0}{2}}a_1
=0,
\end{equation}
with
\begin{equation}
\alpha(\xi)=\frac{C(x)A_0}{\omega L_0},\quad
\beta(\xi)=\frac{D(x)}{\omega L_0^3},\quad
\eta(\xi)=\frac{K(x)}{\omega L_0^{1+\gamma_0}}.
\end{equation}

We choose $A_0$ such that $\alpha(\xi)\equiv 1$.

\subsection*{A.9. Geometric modulation}

Geometry modulates dispersion and damping through
\begin{equation}
\beta(\xi)=\beta_0\big[1+\varepsilon_\beta\cos(q\xi)\big],\qquad
\eta(\xi)=\eta_0\big[1+\varepsilon_\eta\cos(q\xi)\big],
\end{equation}
where $q$ is the geometric wavenumber and $\varepsilon_\beta,\varepsilon_\eta\leq 0.3$.

\subsection*{A.10. Scaling of the dispersion coefficient with Womersley number}

Dimensional consistency of dispersive corrections in oscillatory pipe flow suggests
\[
D\propto \nu^2\mathrm{Wo}^2,
\]
while the pulsation time scale satisfies
\[
T_0=\omega^{-1}=\frac{R_0^2}{\nu \mathrm{Wo}^2}.
\]
Thus
\[
\beta_0=\frac{DT_0}{L_0^3}\propto \frac{\nu R_0^2}{L_0^3}\mathrm{Wo}^{-2}.
\]
We therefore adopt
\begin{equation}
\beta_0 = \beta_{\mathrm{ref}}\,\mathrm{Wo}^{-2},
\end{equation}
consistent with decreasing dispersive coupling at high Womersley number.

Damping is modeled as
\begin{equation}
\eta_0 = \eta_{\mathrm{ref}}\left(1+\frac{0.1}{\mathrm{Wo}}\right).
\end{equation}

The form $\eta_0=\eta_{\mathrm{ref}}(1+C/\mathrm{Wo})$ follows from oscillatory boundary-layer scaling. The Stokes boundary-layer thickness associated with pulsatile motion is $\delta\sim\sqrt{2\nu/\omega}$, hence $\delta/R_0=O(\mathrm{Wo}^{-1})$. Since near-wall shear-driven dissipation scales with the boundary-layer contribution, the leading frequency-dependent correction to an effective damping coefficient is proportional to $\delta/R_0$, motivating $C=O(10^{-1})$ as an order-unity prefactor smaller than one. I choose $C=0.1$ so that this correction is weak in the large-artery regime emphasized in the manuscript (i.e., it remains a small perturbation for $\mathrm{Wo}\gtrsim 5$) while preserving the correct asymptotic trend as $\mathrm{Wo}$ decreases. The constant $0.1$ is therefore a conservative phenomenological coefficient rather than a claimed universal constant.

\subsection*{A.11. Limiting cases}

\begin{itemize}
    \item \textbf{KdV limit}: $\varepsilon_\beta=\varepsilon_\eta=0$ and $\eta_0=0$ gives
    \[
    a_{1,\tau} + a_1 a_{1,\xi} + \beta_0 a_{1,\xi\xi\xi}=0.
    \]

    \item \textbf{Fractional Burgers limit}: $\beta_0=0$ yields
    \[
    a_{1,\tau}+a_1a_{1,\xi}+\eta(\xi)(-\partial_\xi^2)^{(1+\gamma_0)/2}a_1=0.
    \]

    \item \textbf{No geometry limit}: $\varepsilon_\beta=\varepsilon_\eta=0$ reduces the system to a constant-coefficient equation in which spatial parametric forcing is absent. Frequency-dependent optimal spectral redistribution may still occur through the Womersley scaling of $\beta_0$ and $\eta_0$.
\end{itemize}

\subsection*{A.12. Final reduced model}

The final dimensionless model is
\begin{equation}
\boxed{
\frac{\partial a_1}{\partial \tau}
+
a_1\frac{\partial a_1}{\partial \xi}
+
\beta(\xi)\frac{\partial^3 a_1}{\partial \xi^3}
+
\eta(\xi)(-\partial_\xi^2)^{1/2}a_1
=0
}
\end{equation}
with $\beta(\xi)$ and $\eta(\xi)$ given by (A.19). This is the variable-coefficient fractional KdV model governing the amplitude of weakly nonlinear perturbations to pulsatile Womersley flow in geometrically complex arteries. In the weak-modulation, zero-mean case considered in the main manuscript, this variable-coefficient model admits a leading-order homogenized effective-medium limit, described next, which is the form evolved by the numerical solver.

\subsection*{A.13. Homogenized effective-medium limit and numerical implementation}

The spatially modulated coefficients in (A.22) may be decomposed as
\[
\beta(\xi)=\bar{\beta}+\tilde{\beta}(\xi), 
\qquad
\eta(\xi)=\bar{\eta}+\tilde{\eta}(\xi),
\]
where the overbar denotes the spatial mean over one geometric period and 
\(\langle \tilde{\beta} \rangle = \langle \tilde{\eta} \rangle = 0.\)

For the cosine modulation (A.19) with 
\(\varepsilon_\beta,\varepsilon_\eta \le 0.3\),
the fluctuating components satisfy 
\(|\tilde{\beta}|/\bar{\beta}=O(\varepsilon_\beta)\),
\(|\tilde{\eta}|/\bar{\eta}=O(\varepsilon_\eta)\).

Substituting this decomposition into (A.22) gives
\[
a_{1,\tau}
+ a_1 a_{1,\xi}
+ \bar{\beta} a_{1,\xi\xi\xi}
+ \bar{\eta}(-\partial_\xi^2)^{1/2}a_1
=
- \tilde{\beta}(\xi)a_{1,\xi\xi\xi}
- \tilde{\eta}(\xi)(-\partial_\xi^2)^{1/2}a_1.
\]

The right-hand side represents a zero-mean parametric perturbation of order 
\(O(\varepsilon)\).
For weak modulation and long-wave scaling, these terms enter at higher order in the averaged phase-dispersion balance and o not alter the leading-order, spatially averaged dispersion relation that governs the dominant phase-dispersion balance in the weak-modulation regime considered in this work.

Accordingly, the numerical simulations in the main manuscript evolve the homogenized leading-order equation
\[
a_{1,\tau}
+ a_1 a_{1,\xi}
+ \bar{\beta} a_{1,\xi\xi\xi}
+ \bar{\eta}(-\partial_\xi^2)^{1/2}a_1
=0,
\]
which corresponds to the effective-medium limit of the weakly modulated geometry model.

The reported non-monotonic Womersley dependence therefore reflects the frequency-dependent dispersive–dissipative balance of the homogenized operator, while the geometric modulation parameterization specifies the admissible variability bounds and provides the physiological interpretation of the effective coefficients.

\end{document}